\documentclass[%
reprint,
aps,
prl,
amsmath,
amssymb,
superscriptaddress
]{revtex4-2}

\usepackage{graphicx}% Include figure files
\usepackage{dcolumn}% Align table columns on decimal point
\usepackage{bm}% bold math
\usepackage{braket}
\usepackage{scalerel}
\usepackage[hidelinks]{hyperref}% add hypertext capabilities
\allowdisplaybreaks
\usepackage{ lipsum }  
\usepackage{ mathrsfs }
\usepackage{ bbold }
\usepackage{ upgreek }
\usepackage{ dsfont }
\usepackage{ amsmath }
\usepackage{slantsc} % for small capitals
\usepackage{comment}
\usepackage{xr-hyper}  
\usepackage{xcolor}
\usepackage{subfiles}
\usepackage[normalem]{ulem} % \sout{} for strikethrough lines
\newcommand{\Filippos}[1]{\noindent{\textcolor{black}{ #1}}}

\begin{document}
%TC:ignore

\preprint{APS/123-QED}

\title{Dynamical error reshaping for dual-rail erasure qubits}% Force line breaks with \\
% Error-suppressing gates for dual-rail erasure qubits

\author{Filippos Dakis}
 %\email{dakisfilippos@vt.edu}
\affiliation{Department of Physics, Virginia Tech, Blacksburg, Virginia 24061, USA}
\affiliation{Virginia Tech Center for Quantum Information Science and Engineering, Blacksburg, Virginia 24061, USA}
\author{Shruti Puri}
\affiliation{Department of Applied Physics, Yale University, New Haven, Connecticut 06520, USA}
\affiliation{Yale Quantum Institute, Yale University, New Haven, Connecticut 06511, USA}
\author{Sophia E. Economou}
 %\email{economou@vt.edu}
\affiliation{Department of Physics, Virginia Tech, Blacksburg, Virginia 24061, USA}
\affiliation{Virginia Tech Center for Quantum Information Science and Engineering, Blacksburg, Virginia 24061, USA}
\author{Edwin Barnes}
 %\email{efbarne141s@vt.edu}
\affiliation{Department of Physics, Virginia Tech, Blacksburg, Virginia 24061, USA}
\affiliation{Virginia Tech Center for Quantum Information Science and Engineering, Blacksburg, Virginia 24061, USA}
\date{\today}

\date{\today}% It is always \today, today,

\begin{abstract}
Erasure qubits—qubits designed to have an error profile that is dominated by detectable leakage errors—are a promising way to cut down the resources needed for quantum error correction. There have been several recent experiments demonstrating erasure qubits in superconducting quantum processors, most notably the dual-rail qubit defined by the one-photon subspace of two coupled cavities. An outstanding challenge is that the ancillary transmons needed to facilitate erasure checks and two-qubit gates introduce a substantial amount of noise, limiting the benefits of working with erasure-biased qubits. Here, we show how to suppress the adverse effects of transmon-induced noise while performing erasure checks or two-qubit gates. We present control schemes for these operations that suppress erasure check errors by three orders of magnitude and reduce the logical two-qubit gate infidelities by up to three orders of magnitude.
\end{abstract}

\maketitle
%TC:endignore

\begin{figure}[t]
\includegraphics[width=1\linewidth,keepaspectratio]{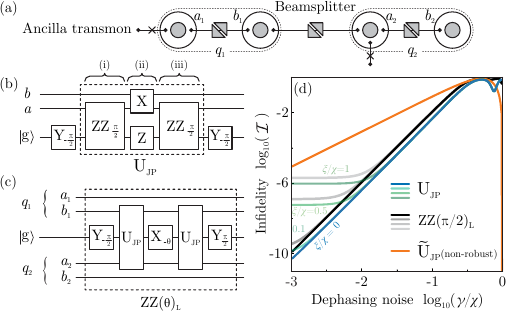}
    \caption{\label{fig:circuits}(a) Schematic of two DR qubits ($q_1$ and $q_2$) coupled by a beamsplitter.  (b)--(c) Circuit diagrams for joint parity and \Filippos{$ZZ(\theta)_{\rm L}$} DCGs. Both diagrams refer to DR qubits, each comprised of two physical cavities $a$ and $b$, while $\ket{\textsl{g}}$ is the ground state of the ancilla qubit. (d) Gate infidelity vs dephasing noise in the ancilla qubit for the robust joint parity gate $U_{\rm JP}$ (blue-green) and logical entangling gate \Filippos{$ZZ(\pi/2)_{\rm L}$} (black-gray). $\xi$ is the dispersive coupling strength in step (ii) of our joint parity check sequence, while $\chi$ is the coupling in steps (i) and (iii). For comparison, the infidelity of the single-shot non-robust joint parity gate $\widetilde{U}_{\rm JP}$ (orange) is also shown.}
\end{figure}

\textbf{\textit{Introduction.}} Meeting the requirements for quantum error correction (QEC) \cite{Knill1998,Preskill1998} and achieving fault-tolerant quantum computing demands substantial improvements in physical qubits and their fundamental operations. Despite rapid progress—from early proposals \cite{Shor1995,Fowler2012,Cochrane1999,Knill2000} to below-threshold demonstrations \cite{Google2024,AWS2024}—practical quantum computing remains constrained by large physical-qubit overhead and the need for complex, high-fidelity multi-qubit operations. Progress therefore hinges on improved hardware, more capable codes, and code–hardware co-design that exploits device-specific noise structure and the natural hierarchy between physical and logical errors \cite{Ruiz2025,Teoh2023}. Error processes are not all equally harmful; detected erasures are comparatively benign, exhibiting higher thresholds and more favorable scaling with code distance than Pauli errors \cite{Gottesman1997,Barrett2010,Grassl1997}.\\
\indent Erasure qubits \cite{Kubica2023}, whose error profiles are dominated by detectable leakage, promise reduced QEC overhead. They have been realized in Rydberg-atom platforms~\cite{Wu2022,Ma2023,Scholl2023}, and more recently in superconducting processors using transmons~\cite{Kubica2023,Levine2024} and microwave cavities~\cite{Chou2024,Teoh2023,Tsunoda2023}. A prominent example is the dual-rail (DR) cavity qubit, which exploits the long coherence times of cavities while encoding a logical qubit in the one-photon subspace of two coupled modes~\cite{Chou2024, Teoh2023}. Photon loss, the dominant error, is converted into a detectable erasure, leaving Pauli errors suppressed by orders of magnitude~\cite{Teoh2023,Chou2024}. However, the transmon ancillae that mediate non-Gaussian operations, such as erasure checks and two-qubit gates, introduce additional noise—ancilla \Filippos{transition frequency fluctuations} and residual $ZZ$ crosstalk \Filippos{originating from an always-on dispersive interaction}—that diminish the erasure bias and limit the performance of DR qubits~\cite{Rosenblum2018,Mundada2019,Ku2020,McKay2019}.\\
\indent In this paper, we show how to substantially improve the performance of DR erasure qubits by dynamically suppressing ancilla-induced errors \Filippos{during erasure checks and two-qubit gates. This requires driving both the DR qubit and the transmon, rendering it a challenging multi-level dynamics problem. A key element of our approach is to simplify the problem by decomposing the dynamics into a sequence of effective two-level systems. The structure of these systems prevents the use of standard dynamical decoupling based on instantaneous pulses, and instead requires finite, continuous waveforms. To construct suitable noise-suppressing pulses, we use} the Space Curve Quantum Control (SCQC) framework \cite{Barnes2022}, in which quantum evolution is mapped to geometric space curves. We derive a geometric condition for $ZZ$ crosstalk cancellation \Filippos{when both qubits are driven simultaneously,} and design two-qubit gates that are robust against both frequency fluctuations and crosstalk noise. The resulting pulses are of low amplitude and narrow bandwidth, making them directly implementable on current platforms. We show that our dynamically corrected gate (DCG) designs substantially improve the \Filippos{quality} of erasure-check gates and logical entangling gates for DR qubits.\\
\indent\textbf{\textit{The system.}} The DR qubit consists of two adjacent microwave cavities coupled via a beam splitter (BS), while a transmon ancilla is dispersively coupled to one of the cavities, see Fig.~\ref{fig:circuits}(a), where two such (coupled) qubits are schematically shown. The logical codewords $\ket{0}_{\rm L}=\ket{10}$ and $\ket{1}_{\rm L}=\ket{01}$ lie in the single-photon manifold, thereby rendering the dominant error---photon loss to state $\ket{00}$---a detectable erasure via measurement of the total photon number parity\Filippos{, which can be facilitated by coupling one cavity to an ancillary transmon (see Fig.~\ref{fig:circuits}(a))} \cite{Tsunoda2023, Teoh2023, Chou2024}. The logical Pauli operators $\{X_1, Y_1,Z_1\}$ are defined as $X_1\equiv \ket{01}\!\bra{10}+\ket{10}\!\bra{01}$, $Y_1\equiv i(\ket{01}\bra{10} - \ket{10}\bra{01})$, and $Z_1 = \ket{10}\bra{10} - \ket{01}\bra{01}$ \cite{SI}. The Hamiltonian of a DR qubit dispersively coupled to an ancillary transmon is \cite{Tsunoda2023}
\begin{equation}
    H_{\rm DR}(t) = \frac{g(t)}{2}\left(e^{i\varphi(t)}a^{\dagger}b + {\rm h.c.}\right) + \delta(t)a^\dagger a - \frac{\chi}{2}a^\dagger a Z_2,
\label{eq:dual_rail_hamiltonian}
\end{equation}
where $g(t)$, $\varphi(t)$ and $\delta(t)$ are the strength, the phase and the detuning of the BS interaction, respectively, $a$ ($a^{\dagger}$) and $b$ ($b^{\dagger}$) are the annihilation (creation) operators acting on the two bosonic modes, $\chi$ is the strength of the dispersive interaction between the ancilla (in the $\textsl{g}$-$f$ manifold) and mode $a$, and  $\{X_2,Y_2,Z_2\}$ are the Pauli matrices in the two-level subspace defined by the ground $\ket{\textsl{g}}$ and second excited $\ket{f}$ levels of the ancilla (e.g. $Z_2 \equiv \ket{\textsl{g}}\bra{\textsl{g}} - \ket{f}\bra{f} $). The intermediate first excited state $\ket{e}$ is reserved for the error-detection of a single ancilla decay event \cite{Rosenblum2018, Reinhold2020}.\\
\indent The dispersive coupling between the DR qubit and the transmon enables one to measure the joint photon number parity of the DR qubit (and hence leakage) as follows. \Filippos{Setting $\delta=0$, $\varphi=0$, $g = \frac{\sqrt{3}}{2}\chi$, and letting the system evolve for $T_g = 2\pi/|\chi|$, the evolution operator becomes \cite{Tsunoda2023,Teoh2023}
\begin{equation}
    \widetilde{U}_{\rm JP} = e^{-i\frac{\pi}{2}N}\left[\mathbb{1}\otimes\ket{\textsl{g}}\bra{\textsl{g}} + e^{i\pi N}\otimes \ket{f}\bra{f}\right]\,,
\label{eq:joint_parity}
\end{equation}
which is the controlled joint parity unitary, with $N = a^\dagger a + b^\dagger b$ the total photon number operator.} This is a single-shot operation that rotates the ancilla about the $z$-axis conditionally on the parity of the total photon number in the cavities. Therefore, by preparing the ancilla in $\ket{+}=(\ket{\textsl{g}}+ \ket{f})/\sqrt{2}$, then applying $\widetilde{U}_{\rm JP}$, and finally performing a \Filippos{$Y(-\pi/2)$ rotation} on the ancilla followed by a measurement, a quantum non-demolition measurement is performed that detects whether an erasure has occurred.\\
\indent \Filippos{Although straightforward, the above erasure-check protocol suffers from a severe drawback:} Equation~\eqref{eq:dual_rail_hamiltonian} does not account for transmon-induced dephasing noise, which can produce an incorrect measurement outcome, leading to an unnecessary reset of the DR qubit state or an undetected erasure. 
\Filippos{A standard metric for capturing parity measurement failures is given by} the average gate fidelity \cite{Nielsen2002, Pedersen2007}.
%\begin{equation}
%    \mathcal{F}(M) = \frac{{\rm tr}(MM^\dagger) + |{\rm tr}(M)|^2}{d(d+1)}\, ,
%\label{eq:Gate_fidelity}
%\end{equation}
%where $M = U_0(T_g)U_g^\dagger$, where $U_0(T_g)$ and $U_g$ are the implemented and target gates, respectively, and $d$ is the Hilbert space dimension. 
%
To test the noise resilience of $\widetilde{U}_{\rm JP}$, we introduce the ancilla \Filippos{frequency fluctuations through a dephasing} noise term $h_n(t)=\frac{\gamma(t)}{2}Z_2$ in Eq.~\eqref{eq:dual_rail_hamiltonian}\Filippos{, with $\gamma(t)$ being a stochastic variable dominated by a $1/f$ power spectral density (PSD) \cite{Burnett2014,Krantz2019, Rower2023, Paladino2014}. For this PSD and for typical gate times in superconducting qubits, the noise can be treated in the quasi-static limit, where $\gamma(t)=\gamma$ is unknown but constant over a single gate or pulse implementation. Note that including the noise directly in the Hamiltonian goes beyond the more common Lindbladian treatment, which presumes memoryless (white) noise, and enables an efficient description of temporally correlated noise such as the $1/f$ noise considered here~\cite{Yan2016, Rower2023, Cywinski2008}.} In Fig.~\ref{fig:circuits}(d) we plot the infidelity, $\mathcal{I} = 1-\mathcal{F}$, of the non-robust joint parity operation $\widetilde{U}_{\rm JP}$ (orange curve) against the dephasing strength. (We reserve the symbol $U_{\rm JP}$ to denote our robust version of this gate described below.) The infidelity scales quadratically, $\mathcal{I}\propto \gamma^2$, confirming the absence of robustness against dephasing noise. This is problematic not only for erasure checks but also for logical entangling gates between DR qubits, which also rely on the joint parity unitary. Figure~\ref{fig:circuits}(c) shows a construction of the logical entangler \Filippos{$ZZ(\theta)_{\rm L}$} from the joint parity unitary and single-qubit rotations \cite{Teoh2023}.\\
\indent\textbf{\textit{Robust joint parity gate.}} Transmon dephasing cannot be suppressed using BS interactions alone (since $[H_{\rm DR},Z_2] =0$); \Filippos{one} must drive the transmon in order to deal with this type of noise. It is challenging to find control pulses that cancel the dephasing noise while implementing the target gate with high fidelity, especially because the inclusion of the transmon drive renders this a multi-level quantum dynamics problem. We circumvent this difficulty by implementing the joint parity unitary as a three-part composite sequence consisting of the following steps: (i) Decouple the cavities and let the transmon state acquire a phase conditional on the number of photons stored in cavity $a$; (ii) Swap the photons between the cavities through the BS interaction while implementing a $Z$ gate on the transmon; (iii) Repeat the first step \Filippos{so that the transmon phase now depends on the total number of photons in both cavities}. These steps are illustrated in Fig.~\ref{fig:circuits}(b). \Filippos{This protocol is similar to the naive one in Eq.~\eqref{eq:joint_parity} in that the dispersive interaction allows the state of the transmon to accumulate information about the photon number parity, while the BS coupling is used to erase ``which-cavity" information in step (ii) so that the logical DR qubit state is not collapsed in the process.} However, dividing up the joint parity operation in this way allows us to reduce the complexity of designing a robust joint parity operation, where we include transmon driving in each step to suppress dephasing; as we show below, in each step of the sequence, the system can be described by time-dependent two-level Hamiltonians, provided we choose the BS coupling and transmon drive parameters appropriately in each step. \Filippos{Note also that the dephasing parameter $\gamma(t)$ is assumed to be approximately constant during each step, but it need not be constant across all three steps.} \\ 
%We then employ the SCQC formalism to design control pulses that implement the target two-level dynamics in each step while suppressing dephasing noise. \\
%
%\indent \Filippos{The transmon is driven via the single-qubit Hamiltonian $H_2(t) = \frac{\Omega_2(t)}{2}\left[\cos\Phi_2(t)X_2 + \sin\Phi_{2}(t)Y_2\right] + \frac{\Delta_2(t)}{2}Z_2$, where $\{\Omega_2, \Phi_2, \Delta_2\}$ are the control fields. The system Hamiltonian then becomes $H(t) = H_{\rm DR}(t) + H_2(t) + h_n(t) \equiv H_0(t) + h_n(t) $, with $H_0(t)$ being the noise-free part.} \\
%
\indent Although the protocol applies to the full two-cavity Hilbert space, for simplicity we focus our analysis on the $\{\ket{00},\ket{01},\ket{10},\ket{11}\}$ subspace. In this truncated space, steps (i) and (iii) are implemented by $ZZ(\pi/2)$ gates that rotate the transmon conditionally on the number of photons in cavity $a$, while in step (ii) an $X_1\otimes Z_2$ gate implements the photon exchange between the cavities and the $Z$ gate on the transmon (see Fig.~\ref{fig:circuits}(b)).  In what follows, we show how to design all these operations to be robust to ancilla dephasing while the intermediate $X_1\otimes Z_2$ operation also cancels the dispersive interaction to first order. The circuit for the joint parity operation is shown in Fig.~\ref{fig:circuits}(b), the net effect of which can be expressed as~\cite{SI}
\begin{equation}
    U_{\rm JP} = \left(P_{00}+P_{11}\right)\otimes\mathbb{1} - X_1\otimes Z_2\,,
\label{eq:joint_parity_Paulis}
\end{equation}
where $P_{ij}=\ket{ij}\bra{ij}$ are the projectors onto two-cavity states $\ket{ij}$. Equations~\eqref{eq:joint_parity}  and \eqref{eq:joint_parity_Paulis} are equivalent up to local rotations, since the former gate in the truncated space reads $\widetilde{U}_{\rm JP} = (P_{00} - P_{11})\otimes \mathbb{1} -i(P_{01} + P_{10})\otimes Z_2$. 

\begin{figure}[t]   
\includegraphics[width=1\linewidth,keepaspectratio]{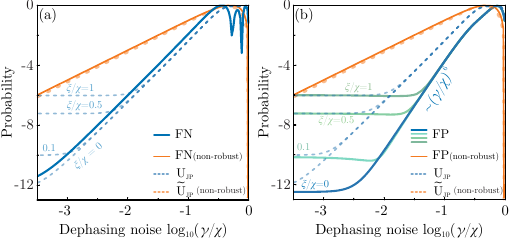}
    \caption{\label{fig:False_negative_main}\Filippos{False erasure assignment probabilities versus noise strength. (a) False negative probability for the robust (blue) and non-robust (orange) erasure-check operations. (b) False positive probability for the robust (blue-green) and non-robust (orange) erasure-check operations. Both axes are logarithmic; for comparison, dashed transparent lines show the infidelity of the joint parity operations. False negative probability does not include the probability of leaking out of the codespace.}}
\end{figure}

\begin{figure}[t]
\includegraphics[width=1\linewidth,keepaspectratio]{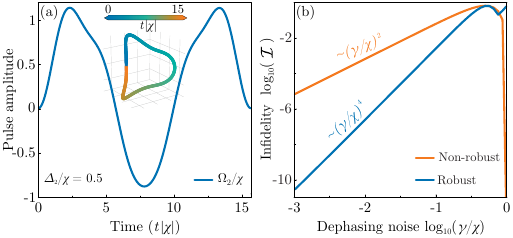}
    \caption{\label{fig:Pulse_and_robustness} Gate design for steps (i) and (iii). (a) Transmon ancilla control field $\Omega_2(t)$ versus time (in units of $1/|\chi|$) for the $ZZ(\pi/2)$ gate obtained from the curvature and the torsion of the curve shown in the inset. \Filippos{For $|\chi|/2\pi = 1\, \rm{MHz}$ \cite{deGraaf2025}, the gate time is $T_{g} \simeq 2.5\, \rm \upmu s$.} (b) Gate infidelity $(\mathcal{I} = 1-\mathcal{F})$ vs dephasing noise strength $(\gamma/\chi)$ showing the robustness against quasi-static dephasing noise to leading-order for the $ZZ(\pi/2)$ gate (blue). For comparison, the infidelity of the non-robust square pulse with amplitude $\Omega/\chi = 5.5\pi$ is also shown (orange).}
\end{figure}

\begin{figure}[t]   
\includegraphics[width=1\linewidth,keepaspectratio]{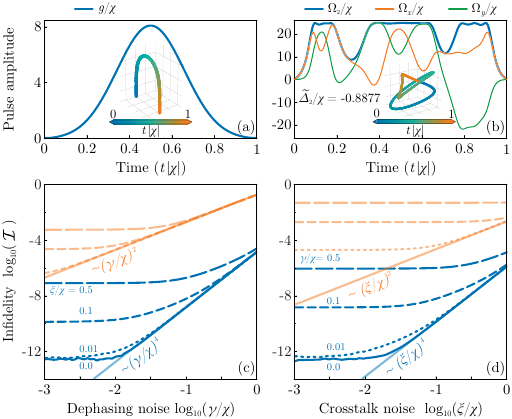}
    \caption{\label{fig:Crosstalk_and_dephasing} Gate design for step (ii). (a)-(b) Control fields versus time (in units of $1/|\chi|$) for the DR and ancilla qubit, respectively, for the $X_1\otimes Z_2$ gate. (c)-(d) Gate infidelity versus dephasing and crosstalk noise for the same gate. (a) BS Gaussian pulse of zero detuning that implements the inter-cavity photon swap. (b) Control fields obtained from the curve shown in the inset using BARQ that implement $Z_2$. For $|\chi|/2\pi = 1\, \rm{MHz}$ \cite{deGraaf2025}, the gate time is $T_{g} = 0.16\, \rm \upmu s$. (c) First-order robustness against $Z_2$ dephasing (blue) for several values of $ZZ$ crosstalk. (d) First-order robustness against $ZZ$ crosstalk (blue) for several values of $Z_2$ dephasing noise. For comparison, the infidelity of non-robust square pulses of constant amplitude $\Omega_1/\chi = 3\pi$ and $\Omega_2/\chi = 9\pi$ is also shown (orange).}
\end{figure}

\Filippos{We first summarize the performance of our robust erasure-check protocol before we elaborate on how we design the transmon drives in each step to achieve this robustness.} The infidelity of the robust joint parity gate ($U_{\rm JP}$) is shown in Fig.~\ref{fig:circuits}(d) to be orders of magnitude smaller than the non-robust single-shot counterpart ($\widetilde{U}_{\rm JP}$) proposed in \cite{Teoh2023,Tsunoda2023}. The dephasing robust gate has quartic sensitivity to noise, $\mathcal{I}\propto\gamma^4$, and saturates only when the parasitic dispersive coupling $\xi$ in step (ii) cannot be eliminated, i.e, $\xi \neq 0$.\\
\indent \Filippos{Average gate fidelity is a faithful metric for quantifying the quality of the implemented operation. For the joint parity operation, however, one is primarily interested in accurately detecting whether the state lies inside or outside the codespace, i.e., the performance for a specific subset of initial and final states. To that end, in Fig.~\ref{fig:False_negative_main} we plot the false negative (FN) and false positive (FP) assignment probabilities. For the non-robust joint parity checks both probabilities scale with $\propto\gamma^2$, showing zero protection against noise. In contrast, for the robust joint parity check the FN probability scales with the fourth power of the noise strength, $P({\rm FN})\propto \gamma^4$, while the FP probability has sextic sensitivity to noise, $P({\rm FP}) \propto \gamma^6$, revealing thus first- and second-order dephasing noise suppression, respectively \cite{SI}. The two misassignments differ in both frequency and severity. False positives are the more common failure, and while they increase the effective erasure rate, they do not produce logical errors~\cite{Teoh2023, Chou2024}. False negatives are second-order events, requiring both a leakage event with probability $p_{\rm erasure}$ and a failure of the erasure check to detect it, but the resulting undetected leakage can propagate into logical errors, which are the most damaging in a stabilizer code~\cite{Chou2024}. The asymmetric cancellation of ancilla dephasing noise in Fig.~\ref{fig:False_negative_main} works in our favor, suppressing FP to second order and the already rare FN to first order~\cite{SI}. The same protocol can be applied to the DR configuration with two ancillae \cite{Chou2024}, with similar performance expected.}\\
\indent \Filippos{We now explain how we design the transmon drives in each step of the robust erasure-check protocol.} \Filippos{We begin by projecting $H_{\rm DR}(t)$ onto the single-photon subspace~\cite{SI}, where for $\varphi(t)=\delta(t)=0$ the noiseless dynamics are governed by
\begin{equation}
    H_0(t) = \frac{g(t)}{2}X_1 - \frac{\chi}{4}Z_1\otimes Z_2
    -\frac{\chi}{4}Z_2 + H_2(t)\, ,
\label{eq:two_qubit_hamiltonian}
\end{equation}
with $\{X_1,Z_1\}$ acting on the DR qubit, and $H_2(t) = \frac{\Omega_2(t)}{2}\left[\cos\Phi_2(t)X_2 + \sin\Phi_{2}(t)Y_2\right] + \frac{\Delta_2(t)}{2}Z_2$ being the ancilla drive Hamiltonian. For the $ZZ(\pi/2)$ gates of steps (i) and (iii) [Fig.~\ref{fig:circuits}(b)] we switch off the BS interaction ($g=0$) and set $\Delta_2=\chi/2$, so that the Hamiltonian is block diagonal even when we include dephasing: $H^{ZZ}=\ket{0}\bra{0}\otimes H_{-}+\ket{1}\bra{1}\otimes H_{+}$ with $H_{\pm}=\frac{\Omega_2(t)}{2}X_2\pm\frac{\chi}{4}Z_2+\frac{\gamma}{2}Z_2$. The two-qubit problem collapses onto a pair of single-qubit problems that differ only in the sign of the dispersive shift, and it is this structure that we exploit. \Filippos{Here, it is evident that we cannot use instantaneous dynamical decoupling pulses to cancel the dephasing term, as any such sequence would also cancel the dispersive coupling, which is critical for acquiring the desired parity information. The two $Z_2$ terms can only be distinguished using finite, continuous pulses, and hence we turn to SCQC for this task \cite{SI}.} In the SCQC formalism each block maps onto a space curve whose curvature is set by the drive amplitude and whose torsion is set by the detuning, and dephasing is cancelled to first order precisely when the curve closes~\cite{Zeng2019, SI}. What we require is therefore a single closed curve of constant torsion that implements a $Z(\pi/2)= e^{i\frac{\pi}{4}Z}$ rotation. We obtain such a curve from the analytic family of constant-torsion closed curves of Ref.~\cite{Calini1996}, whose curvature is an elliptic cosine and is smooth enough to be realized as a pulse; tuning the torsion and the Fourier components of the curvature yields $U_{+}(T_g)=Z(\pi/2)$ to infidelity $10^{-14}$~\cite{SI}. Because the two blocks share the same curvature but opposite torsion, the same pulse produces $U_{-}(T_g)=Z(-\pi/2)$~\cite{SI}, and hence both blocks together yield the desired entangling gate $ZZ(\pi/2)$ that imprints the photon number parity onto a relative phase in the transmon state. Figure~\ref{fig:Pulse_and_robustness} shows the curve, the resulting pulse envelope given by the curvature, $\Omega_2(t)=\kappa(t)$, and the gate infidelity, revealing that the corrected gate scales as $\gamma^4$ rather than the $\gamma^2$ scaling of the non-robust gate.}\\
\indent\Filippos{In many architectures the dispersive coupling $\chi$ is always-on and not tunable, so the ancilla-cavity interaction remains on during single-qubit gates; even when it is tunable~\cite{Maiti2025}, a finite on-off ratio leaves residual $ZZ$ crosstalk that limits gate time and fidelity. Therefore, for step (ii), we treat this coupling as noise rather than as a resource; to emphasize this we rewrite Eq.~\eqref{eq:two_qubit_hamiltonian} as $H^{XZ}(t) = H_0^{XZ}(t) + \tilde{h}_n$ with $H_0^{XZ}(t) = \frac{g(t)}{2}X_1 + \tilde{H}_2(t)$, $\tilde{H}_2(t) = \frac{\Omega_2(t)}{2}\left[\cos\Phi_2(t)X_2 + \sin\Phi_{2}(t)Y_2\right] + \frac{\tilde{\Delta}_2(t)}{2}Z_2$ and $\tilde{\Delta}_2(t) = \Delta_2(t) - \frac{\chi}{2}$, and $\tilde{h}_n = \frac{\gamma}{2}Z_2 - \frac{\xi}{4}Z_1\otimes Z_2$, where both $\gamma$ and $\xi$ are quasi-static~\cite{SI}. The residual coupling $\xi$ is left independent of $\chi$ to accommodate  stochastic or imperfectly cancelled interactions. To leading order in the Magnus expansion~\cite{Blanes2009}, the crosstalk contributes~\cite{SI}
\begin{equation}
    \int_{0}^{T_g} dt\, (U_{1}^{\dagger}Z_1U_1)\otimes(U_{2}^{\dagger}Z_2U_2)
    = \vec{\sigma}_1^{T}\mathcal{M}\vec{\sigma}_2,
\label{eq:ZZ_integral_SCQC}
\end{equation}
with $U_i$ generated by the single-qubit Hamiltonians and $\vec{\sigma}_i=[X_i,Y_i,Z_i]$. Within SCQC each factor is the tangent vector of the corresponding space curve, $\vec{T}_i$, so the coefficient matrix factorizes into an outer product and the cancellation condition becomes~\cite{SI}
\begin{equation}
    \mathcal{M}_{ij}= \int_{0}^{T_g} dt\, T_{1}^{i}(t)T_{2}^{j}(t)  = 0\, .
\label{eq:crosstalk_condition}
\end{equation}
Equation~\eqref{eq:crosstalk_condition} states that quasi-static $ZZ$ noise cancels when the components of the two tangent vectors are orthogonal functions on $[0,T_g]$. Note that the $ZZ$ suppression condition is compatible with the cancellation conditions for other types of noise such as dephasing or pulse amplitude noise~\cite{Nelson2023}, it just adds an extra global geometric constraint. Moreover, Eq.~\eqref{eq:crosstalk_condition} is general in the sense that it holds when both qubits are driven in parallel while allowing for arbitrary rotations $e^{i\frac{\theta_1}{2}\mathbf{\hat{n}}_1\cdot\vec{\sigma}}\otimes e^{i\frac{\theta_2}{2}\mathbf{\hat{n}}_2\cdot\vec{\sigma}}$, features that existing crosstalk cancellation schemes \cite{Watanabe2024, ORBIT2014, Yi2024, Sheldon2016} do not offer.} \Filippos{In the Supplemental Information~\cite{SI}, we also discuss alternative approaches. For instance, spin-echo-type methods can suppress the crosstalk, but they require at least four echo pulses, which leads to a large pulse overhead and thus longer gate times. We further show that, when only one qubit is driven and the other is idle, Eq.~\eqref{eq:crosstalk_condition} reduces to the closed-curve condition.}\\
\indent\Filippos{With the crosstalk cancellation condition in hand, we design a robust version of the step (ii) operation $U_0 = X_1\otimes Z_2$, which swaps the photon between cavities while flipping the ancilla phase [Fig.~\ref{fig:circuits}(b)]. We drive the BS with a Gaussian waveform along the $x$-axis, giving the closed-form tangent vector $\vec{T}_1(t)=\cos\!\big(\!\int_0^t\!\kappa_1\big)\hat{x} +\sin\!\big(\!\int_0^t\!\kappa_1\big)\hat{y}$ with Gaussian $\kappa_1(t)$~\cite{SI}, and place the entire robustness burden (dephasing and crosstalk) on the ancilla curve. For that curve, we use the Bézier ansatz for robust quantum control (BARQ)~\cite{Piliouras2026}, which constructs SCQC curves from optimized control points and, crucially, satisfies both the closure condition and the target gate by construction. Numerical optimization of the curve is therefore only required to enforce Eq.~\eqref{eq:crosstalk_condition}, where $\vec{T}_1$ is fixed as above, substantially simplifying the optimization. Extracting the driving fields from the optimized transmon curve yields a gate robust to first order in both $\gamma$ and $\xi$. Figure~\ref{fig:Crosstalk_and_dephasing} shows the resulting control fields and the infidelity across the noise plane; both pulses have low peak amplitude and narrow bandwidth.}\\
\indent\textbf{\textit{Robust logical gates.}} In the DR architecture, logical entangling gates are realized by combining the joint parity unitary $U_{\rm JP}$ with ancilla rotations [see Fig.~\ref{fig:circuits}(c)]. Hardware-wise, this requires only an additional coupler between two neighboring cavities, $b_1$ and $a_2$, which are part of the discrete logical qubits $q_1$ and $q_2$, respectively [see Fig.~\ref{fig:circuits}(a)]. Instead of applying $U_{\rm JP}$ between the cavities of $q_1$ and the ancilla [see Fig.~\ref{fig:circuits}(b)], we apply this gate to cavities $b_1$ and $a_2$ together with the transmon [see Fig.~\ref{fig:circuits}(c)]. Importantly, applying the robust joint parity gate derived in Eq.~\eqref{eq:joint_parity_Paulis} on cavities of different logical qubits (control and target) \Filippos{returns the state in the} subspace spanned by $\{\ket{00},\ket{01},\ket{10},\ket{11}\}$. As an example, we consider the \Filippos{$ZZ(\theta)_{\rm L}$} gate, which is a perfect entangler for $\theta =\pi/2$ and locally equivalent to \textsc{cz} and \textsc{cnot}. The \Filippos{$ZZ(\theta)_{\rm L}$} gate is implemented when the robust joint parity operation is applied twice, interleaved by a single-qubit $X(-\theta)$ rotation on the ancilla \cite{SI}. The circuit diagram of the gate and the gate infidelity, in the case of $\theta=\pi/2$, are shown in Fig.~\ref{fig:circuits}(c)-(d), respectively. As expected, the gate infidelity follows the joint parity slope but is shifted upward, reflecting extra noise accumulation from the longer gate time. Also, crosstalk noise has the same effect on \Filippos{$ZZ(\pi/2)_{\rm L}$} as on $U_{\rm JP}$. \\
\indent\textbf{\textit{Conclusion and outlook.}} We showed how to improve the performance of erasure checks and logical two-qubit gates in dual-rail erasure qubits based on superconducting cavities by \Filippos{suppressing transmon dephasing noise during these operations. This was facilitated by decomposing the dynamics into simple effective two-level systems. Because the structure of these systems requires finite, continuous pulses for noise cancellation, we employed the SCQC formalism to design smooth} pulses that dynamically suppress both quasi-static dephasing noise and $ZZ$ crosstalk. Our methods reduce erasure-check errors and logical two-qubit gate infidelities by three orders of magnitude, as is evident from Fig.~\ref{fig:circuits}(d) for typical levels of dephasing noise $\gamma/\chi\sim10^{-2}$~\cite{Chou2024,Teoh2023}. By making the dispersive coupling tunable~\cite{Maiti2025}, we not only anticipate the gate infidelities to reduce by an additional order of magnitude, but that this can also be achieved with shorter gate times and smoother waveforms.\\ 
\indent In addition to having higher fidelities, our gates also lower calibration requirements by being insensitive to slow drifts of transmon frequencies. Moreover, preserving the erasure bias during parity checks and two-qubit gates can accelerate decoding and reduce QEC overhead. In future work, it would be valuable to quantify the QEC gains enabled by our gates.\\
%
%TC:ignore
\indent The scripts used for the simulations and the corresponding data are available at \cite{Dakis_DR_github}.\\
\indent\textit{Acknowledgments.} We thank Steve Girvin and Evangelos Piliouras for valuable discussions. EB acknowledges support from the National Science Foundation (grant no. 2137776) and the Office of Naval Research (grant no. N00014-25-1-2125). SEE and SP acknowledge support from the U.S. Department of Energy, Office of Science, National Quantum Information Science Research Centers, Co-design Center of Quantum Advantage (C2QA) under Contract No. DE-SC0012704. \\

\bibliographystyle{apsrev4-2}
\makeatletter
\let\oldaddcontentsline\addcontentsline
\let\addcontentsline\@gobblethree
\bibliography{apssamp}
\let\addcontentsline\oldaddcontentsline
\makeatother

%TC:ignore
\clearpage

\onecolumngrid

\begin{center}
    \vspace*{0.5em}
    {\Large \textbf{Supplemental Information}}
\end{center}

% --- S-numbering (do this after the heading)
\setcounter{equation}{0}\renewcommand{\theequation}{S\arabic{equation}}
\setcounter{figure}{0}\renewcommand{\thefigure}{S\arabic{figure}}
\setcounter{table}{0}\renewcommand{\thetable}{S\arabic{table}}

\Filippos{
\addtocontents{toc}{\protect\setcounter{tocdepth}{2}} % re-enable from here on
\tableofcontents
\vspace{1em}
}
\setcounter{secnumdepth}{3}
\setcounter{section}{0}
\renewcommand{\thesection}{\Roman{section}}
\renewcommand{\thesubsection}{\thesection.\Alph{subsection}}
\makeatletter
\renewcommand{\p@subsection}{}
\makeatother
\section{Space Curve Quantum Control for single-qubit gates \label{sec:single-qubit scqc} }
The Space Curve Quantum Control (SCQC) framework \cite{Zeng2019,Barnes2022} exploits the correspondence between quantum evolution and geometric space curves. This approach provides a global view of control fields that realize a target gate while satisfying noise-cancellation conditions, with clear geometric interpretations. For instance, closed curves cancel first-order dephasing and closed curves with zero signed area cancel this noise up to second order \cite{Zeng2019}, while curves whose tangent vectors sweep out zero area when projected onto any plane cancel multiplicative error on the driving field envelope \cite{Nelson2023}. Finally, the implemented gate is determined by the relative orientation of the Frenet–Serret frame between the initial and final points of the curve.\\
\subsection{Dephasing robust condition}
\indent Although SCQC can be applied to systems with any finite number of Hilbert space dimensions \cite{Buterakos2021}, here we focus on the case of a two-level system. We start by considering the noise-free single-qubit Hamiltonian that governs a two-level system in the lab frame ($\hbar = 1$)
\begin{equation}
    \mathcal{H}_{\rm lab}(t) = -\frac{\omega_q}{2}Z + \Omega(t)\cos(\omega_d t -\Phi(t))X\, ,
\label{eq:hamiltonian_labframe}
\end{equation}
where $X, Y, Z$ are the Pauli matrices, $\omega_q$ is the qubit transition frequency for the $\ket{0}\leftrightarrow\ket{1}$ transition, $\omega_d$ is the carrier frequency of the driving field, $\Omega(t)$ is the envelope of the driving field (Rabi rate), and $\Phi(t)$ is its phase. By moving to the frame rotating with the driving frequency defined by $U_d = e^{i\frac{\omega_d}{2}tZ}$ and invoking the rotating-wave approximation (RWA), to neglect the fast oscillating terms, we obtain the effective, noise-free Hamiltonian 
\begin{equation}
    \mathcal{H}_0(t) = \frac{\Omega(t)}{2}\left[\cos\Phi(t)X + \sin\Phi(t)Y\right] + \frac{\Delta}{2}Z \, ,
\label{eq:hamiltonian_single-qubit}
\end{equation}
where $\Delta = \omega_d - \omega_q$ is the detuning. Hamiltonian $\mathcal{H}_0$ is the point of reference for the SCQC analysis that follows below. Now that the noiseless Hamiltonian is defined, we can proceed by introducing environmental noise in the qubit transition frequency. This type of noise can conveniently be modeled as an additive longitudinal error $\gamma$ in $\Delta$, leading to the noisy Hamiltonian:
\begin{equation}
    \mathcal{H}(t) = \mathcal{H}_0(t) + h_n(t) = \frac{\Omega(t)}{2}\left[\cos\Phi(t)X + \sin\Phi(t)Y \right] + \frac{\Delta(t) + \gamma(t)}{2}Z \, .
\label{eq:Hamiltonian_noisy}
\end{equation}
In general, $\gamma(t)$ is a stochastic process; however, here we focus on the quasi-static limit where $\gamma(t) = \gamma$ is unknown but constant over a gate implementation (i.e., it varies on timescales much longer than the gate time). In superconducting devices, such low-frequency shifts are commonly attributed to qubit miscalibrations \cite{Amer2025} and flux noise \cite{Krantz2019}, often summarized phenomenologically as $1/f$ noise \cite{Krantz2019}; the quasi-static approximation captures their dominant impact during a single gate. All the theoretical derivations discussed here are covered extensively in \cite{Zeng2019, Nelson2023, Piliouras2026}. \\
\indent We can isolate the effect of this noise term on the evolution by moving to the interaction frame. This is done using the decomposition of the total evolution operator as $U = U_0U_I$ and then solving the effective Schr\"{o}dinger equation for $U_I$ :
\begin{equation}
    \dot{U}_I=-i(U_0^\dagger h_n U_0)U_I = -i \tilde{h}_nU_I\,,
\label{eq:schrodinger_eq_noise_operator}
\end{equation}
where $U_0(t) = \mathcal{T}{\rm exp}\{-i\int_0^tdt'\,\mathcal{H}_0(t')\}$ is the noiseless evolution operator, and $U_I(t) = \mathcal{T}{\rm exp}\{-i\int_0^tdt'\,\tilde{h}_n(t')\}$ is the noise evolution operator ($\mathcal{T}$ denotes the time ordering operator). The Magnus expansion \cite{Blanes2009} of the interaction frame evolution operator is then controlled by the small parameter $\gamma$. At first order we have:
\begin{equation}
    U_I(t) \approx e^{-i\Pi_1(t)}\, ,
\label{eq:magnus_first_order}
\end{equation}
where
\begin{equation}\label{eq:first_order_error}
    \Pi_1(t) = \int_0^t dt'\, \tilde{h}_n(t') = \frac{\gamma}{2}\int_0^t dt' U_0^\dagger(t')Z U_0(t')\,.
\end{equation}
We can suppress the errors induced by noise $\gamma$ in $U_I$ at the gate time $T_g$, which marks the end of the evolution, through the following condition:
\begin{equation}
    \int_0^{T_g} dt\,U_0^\dagger ZU_0 = 0\, ,
\label{eq:dephasing_condition}
\end{equation}
which ensures that $\Pi_1$ vanishes and therefore $U_I(T_g) \approx 1$. The correspondence to differential geometry can be established by using Eq. \eqref{eq:first_order_error} to define a \emph{space curve} or \emph{error curve} as in Ref.~\cite{Zeng2019}:
\begin{equation}
     \int_0^{t} dt\,U_0^\dagger ZU_0 = \vec{r}(t)\cdot \vec{\sigma}\, ,
\label{eq:error_curve}
\end{equation}
where $\vec{r}(t)$ is the position vector of a three-dimensional curve, and $\vec{\sigma} = [X, Y, Z]^T$ is the Pauli vector. This definition maps the dephasing robustness condition in Eq. \eqref{eq:dephasing_condition} to the analytic geometric closed curve condition: 
\begin{equation}
    \vec{r}(T_g) = \vec{r}(0)\, .
\label{eq:closed_curve_condition}
\end{equation}
This definition lies at the core of the SCQC formalism and associates dephasing robust controls with closed loops in 3D, where time is equivalent to the length along the curve.\\
\indent To study these 3D space curves, we define an orthonormal frame called the Frenet-Serret frame \cite{Pressley2010}, consisting of 
\begin{equation}
    \vec{T}\equiv \dot{\vec{r}}\,,\qquad \vec{N} \equiv \dot{\vec{T}}/\|\dot{\vec{T}}\|\,,\qquad \vec{B} \equiv \vec{T}\times \vec{N}\,,
\label{eq:tangent_normal_binormal}
\end{equation}
which are the tangent, normal, and binormal vectors, respectively, $\|\vec{A}\| = \sqrt{A_1^2 + A_2^2 + A_3^2}$ is the Euclidean norm, and $\dot{\vec{A}}$ denotes the derivative with respect to time. These vectors then satisfy the Frenet-Serret equations,
\begin{equation}
\dot{\vec{T}} = \kappa\vec{N}\,,\qquad \dot{\vec{N}}= -\kappa \vec{T} + \tau\vec{B}\,,\qquad \dot{\vec{B}} = -\tau \vec{N}.
\label{eq:frenet_serret}
\end{equation}
Once we find a closed space curve, we can find the corresponding control fields $\Omega$, $\Phi$, and $\Delta$ of Eq. \eqref{eq:hamiltonian_single-qubit} from the curvature $\kappa$ and torsion $\tau$ of the space curve:
\begin{equation}
    \Omega(t) = \kappa(t) = \dot{\vec{T}}\cdot\vec{N}\,,\qquad \dot{\Phi}(t)-\Delta = \tau(t) = \dot{\vec{N}}\cdot\vec{B}\,.
\label{eq:fields_curvature_torsion}
\end{equation}
Note that the detuning $\Delta$ and the phase field $\Phi$ are not uniquely determined by the geometry of the space curve. Geometrically, the curvature quantifies how much the curve bends (deviates from a straight line), and the torsion quantifies how much the curve twists (deviates from a planar curve). We also note that we use a slightly modified Frenet-Serret frame compared to the one summarized above to allow for a signed curvature $\kappa(t)$ that is smooth and can assume both positive and negative values as $t$ is varied.\\
\indent Of course, the SCQC approach extends to higher orders of dephasing noise cancellation \cite{Zeng2019}, and different noise types such as multiplicative error on the driving fields \cite{Piliouras2026} and crosstalk noise shown in this paper.\\
\subsection{Adjoint representation}
\indent Next, we present the adjoint representation \cite{Byrd2002, Piliouras2026} to streamline the process of gate-fixing in SCQC. By definition, for a unitary $U$, the elements of the adjoint representation of the evolution are given by
\begin{equation}
    \mathcal{R}_{U}^{ij} = \frac{1}{2}{\rm tr}(U^\dagger \sigma_i U\sigma_j)\, .
\end{equation}
In SCQC the implemented gate $U_0$ is given by 
\begin{equation}
    \mathcal{R}_{U_{0}}(t) = \mathcal{R}_Z(\Phi(t))\mathcal{R}_F(t)\mathcal{R}_F^T(0)\, ,
\label{eq:adjoint_representation}
\end{equation}
where the rows of the matrix $\mathcal{R}_F$ contain the Frenet-Serret vectors
\begin{equation}
    \mathcal{R}_F(t) = \left[-\vec{B}, \vec{N}, \vec{T}\right]^T\, ,    
\label{eq:Rf_matrix}
\end{equation}
while
\begin{equation}
    \mathcal{R}_Z(\Phi(t)) = 
                        \begin{bmatrix}
                            \cos\Phi(t) & -\sin\Phi(t) & 0\\
                            \sin\Phi(t) & \hphantom{-}\cos\Phi(t) & 0 \\
                            0 & 0 & 1
                        \end{bmatrix}    
\end{equation}
is a rotation by angle $\Phi(t)$ about the $z$-axis. For time $t=0$, Eq. \eqref{eq:adjoint_representation} returns $\mathcal{R}_{U_0}(0) = {\rm  diag}(1,1,1)$ which translates to the identity unitary $U_0(0) = I$ on the qubit.
\subsection{\label{subsec:negative_torsion}Curves of same curvature and different torsion}
\noindent Consider two closed curves $\vec{\alpha}(t)$ and $\vec{\beta}(t):=P\vec{\alpha}(t)$ (i.e., $\vec{\beta}$ is the transformation of $\vec{\alpha}$ under parity or reflection, ${\rm det}(P) = -1$) where their frame vectors $[\vec{T},\vec{N},\vec{B}]$ satisfy the following equations:
\begin{equation}
    \vec{T}_{\beta} = P\vec{T}_{\alpha},\qquad \vec{N}_{\beta} = P\vec{N}_{\alpha}\,,\qquad \vec{B}_{\beta} = -P\vec{B}_{\alpha}.
\end{equation}
The curvature is a scalar and remains unaffected, while the torsion is a pseudo-scalar and thus changes sign:
\begin{equation}
\begin{split}
    \kappa_{\beta} &=  \dot{\vec{T}}_\beta\cdot \vec{N}_\beta = (P\dot{\vec{T}}_\alpha)\cdot (P\vec{N}_\alpha) = (P\dot{\vec{T}}_\alpha)^{T}(P\vec{N}_\alpha) = \dot{\vec{T}}_\alpha^T P^T P \vec{N}_\alpha) = \dot{\vec{T}}_\alpha^T \vec{N}_\alpha = \kappa_\alpha\, , \\
    &\phantom{=}\\
    \tau_\beta &= - \vec{N}_\beta \cdot\dot{\vec{B}}_\beta = - (P\vec{N}_\alpha) \cdot(-P\dot{\vec{B}}_\alpha) = \vec{N}_\alpha^{T}\dot{\vec{B}}_\alpha = - \tau_\alpha\, .
\end{split}
\end{equation}
Now assume that curve $\vec{\alpha}$ transforms under parity ($P = {\rm diag}(-1,-1,-1)$), or under reflection in the $yz-$plane ($P = {\rm diag}(-1, 1, 1)$) and that $\vec{\alpha}$ implements the rotation $\mathcal{R}_{U_\alpha} = \mathcal{R}_{F_\alpha}(t)\mathcal{R}_{F_\alpha}^T(0)$, where we set the phase $\Phi(t) =0$ as we do in the main text, then $\vec{\beta}$ implements 
\begin{equation}
\begin{split}
    \mathcal{R}_{U_\beta} =\mathcal{R}_Z(\Phi(t))\mathcal{R}_{F_\beta}(t)\mathcal{R}_{F_\beta}^T(0) &= \mathcal{R}_Z(\Phi(t)) R_X(\pi)\mathcal{R}_{F_\alpha}(t)\mathcal{R}_{F_\alpha}^T(0)R_X(\pi)\\
    &\phantom{=}\\
    &\overset{\Phi =0 }{=} R_X(\pi)\mathcal{R}_{F_\alpha}(t)\mathcal{R}_{F_\alpha}^T(0)R_X(\pi) = R_X(\pi)\mathcal{R}_{U_\alpha}(t)R_X(\pi)\, .
\end{split}
\end{equation}
We see that in the adjoint representation the rotation implemented due to curve $\vec{\beta}$ is $R_{U_\beta}(t)=R_X(\pi)\mathcal{R}_{U_\alpha}(t)R_X(\pi)$, which translates to the unitary $U_{\beta}(t) = XU_{\alpha}(t)X$. Hence, by choosing $U_{\alpha}(T_g) = Z(\theta)$ we obtain $U_{\beta}(T_g) = XZ(\theta)X = Z(-\theta)$. Thus, we see that curves with the same curvature and torsions that differ only by a sign lead to the same gate conjugated by $X$ rotations.
\Filippos{
\subsection{Pulse construction for the three steps of the parity check \label{subsec:pulse_construction}}
\noindent Here we provide additional information for the curves and the control fields used in the three-part composite sequence of the joint parity gate.\\
\indent For steps (i) and (iii) [Fig.~\ref{fig:circuits}(b)] we set $g=0$ and $\Phi_2 = 0$ in Eq.~\eqref{eq:two_qubit_hamiltonian}, and choose $\Delta_2 = \chi/2$ to cancel the bare ancilla shift $-\frac{\chi}{4}Z_2$. This leaves $H_{0}^{ZZ}(t) = -\frac{\chi}{4}Z_1\otimes Z_2 + \frac{\Omega_2(t)}{2}X_2$, which together with $h_n = \frac{\gamma}{2}Z_2$ yields the block-diagonal form quoted in the main text. It therefore suffices to design a single curve, since the two blocks differ only in the sign of $\chi/4$. Within SCQC the constant torsion $\tau$ and the time-dependent curvature $\kappa(t)$ act as control knobs for a target rotation $U_g = Z(\theta) = e^{i\frac{\theta}{2}Z}$, while closure of the curve enforces first-order dephasing cancellation~\cite{Zeng2019}. The closed constant-torsion curves of Ref.~\cite{Calini1996} satisfy both requirements. They are parametrized by a geometric modulus $p\in[0,1]$, with curvature given by the elliptic cosine $\kappa(x) = {\rm cn}(x,p)$ and torsion $\tau = \frac{1}{2p}(2E/K-1)^{1/2}$, where $K\equiv K(p)$ and $E\equiv E(p)$ are the complete elliptic integrals of the first and second kind. Since $\Omega_2(t)$ is directly proportional to $\kappa(t)$, the smoothness of the elliptic cosine is what makes the resulting pulse experimentally friendly. For reference, for $p=0$ the elliptic cosine reduces to the trigonometric cosine, i.e., ${\rm cn}(x,0) = \cos(x)$.\\
\indent We construct the curve for the noise-free part of $H_{+}$ as follows. As an ansatz we take $p = 0.63093$, yielding a closed curve of constant torsion $\tau = 0.6018$~\cite{Calini1996} that generates $Z(\pi)$. Adjusting $\tau$ and the Fourier components of $\kappa(t)$ then gives $U_{+}(T_g) = Z(\pi/2)$ with infidelity $\mathcal{I} = 1-\mathcal{F} = 10^{-14}$. Because $H_{-}$ evolves under the same curvature with opposite torsion, its gate is $U_{-}(T_g) = XZ(\pi/2)X = Z(-\pi/2)$, and the two blocks combine into $U_{0}^{ZZ}(T_g) = \ket{0}\bra{0}\otimes U_{-} + \ket{1}\bra{1}\otimes U_{+} = ZZ(\pi/2)$, realized with $\Omega_2(t) = \kappa(t)$, $\Delta_2 = \chi/2$, and $T_g = 15.7/|\chi|$. The curve, Rabi rate, and the robustness against dephasing noise are shown in Fig.~\ref{fig:Pulse_and_robustness}. Similarly, alternative curves follow from other initial values of $p$ as shown in Sec.~\ref{sec:additional_results}.\\
\indent For step (ii) we regard the dispersive coupling in Eq.~\eqref{eq:two_qubit_hamiltonian} as an unwanted interaction, so that the noise-free Hamiltonian is $H_0^{XZ}(t) = H_1(t) + \tilde{H}_2(t)$ with $H_1(t) = \frac{g(t)}{2}X_1$, $\tilde{H}_2(t) = \frac{\Omega_2(t)}{2}\left[\cos\Phi_2(t)X_2 + \sin\Phi_{2}(t)Y_2\right] + \frac{\tilde{\Delta}_2(t)}{2}Z_2$, and $\tilde{\Delta}_2(t) = \Delta_2(t) - \frac{\chi}{2}$. The crosstalk is absorbed into the noise Hamiltonian $\tilde{h}_n = \frac{\gamma}{2}Z_2 - \frac{\xi}{4}Z_1\otimes Z_2$, where $\chi$ has been replaced by $\xi$ to allow the residual interaction to deviate from the bare dispersive shift, whether through an additional stochastic coupling between the qubits or through imperfect tuning. The evolution operators $U_1$ and $U_2$ entering Eq.~\eqref{eq:ZZ_integral_SCQC} are generated by $H_1$ and $\tilde{H}_2$, respectively, and the tangent vectors are defined by $U_i^{\dagger}Z_iU_i = \vec{T}_i\cdot\vec{\sigma}_i$, with $T_i^{k}(t)$ the component along the $k$ axis, $k\in\{x,y,z\}$.\\
\indent We set the gate time to $T_g = 1$ and drive the DR qubit along the $x$ axis only ($\varphi = \delta = 0$) with $g(t) = \kappa_1(t) = A_0\exp\!\big[\!-(t-t_0)^2/\alpha^2\big]$ and $A_0 = 5\pi/[{\rm erf}(5/2)\sqrt{\pi}]$, $t_0 = 1/2$, $\alpha = 0.2$, implementing $X_1$. The associated tangent vector follows in closed form, $\vec{T}_1(t) = \cos\!\big(\!\int_0^t\!dt'\,\kappa_1\big)\hat{x} + \sin\!\big(\!\int_0^t\!dt'\,\kappa_1\big)\hat{y}$. For the ancilla we use BARQ~\cite{Piliouras2026}, which builds SCQC curves from Bézier control points and satisfies the closure condition and the target gate by construction, so that optimization is needed only for additional robustness constraints. These may be taken to be the curve's signed area (second-order dephasing cancellation), the signed area of the tangent curve (first-order multiplicative-noise cancellation)~\cite{Nelson2023}, or the orthogonality conditions of Eq.~\eqref{eq:crosstalk_condition}. Fixing $U_g = Z_2$ upfront and setting $\vec{T}_1(t)$ to the expression above, we optimize $\vec{T}_2(t)$ until Eq.~\eqref{eq:crosstalk_condition} is satisfied and extract the ancilla drive $\Omega_2(t)$ from the resulting curve. The curves, the control fields, and the robustness against dephasing and crosstalk noise are shown in Fig.~\ref{fig:Crosstalk_and_dephasing}.}
\section{Crosstalk cancellation condition \label{sec:crosstalk_condition}}
\begin{figure}[b!]
\includegraphics[width=1\linewidth,keepaspectratio]{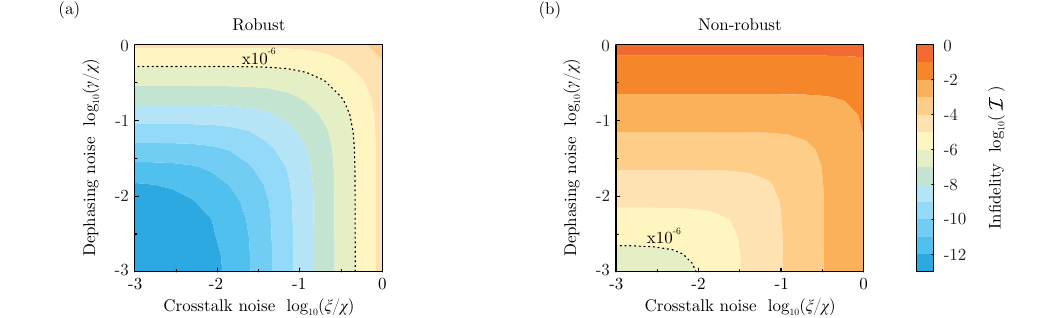}
    \caption{\label{fig:Contour_plot} Gate infidelity ($\mathcal{I} = 1-\mathcal{F}$) for an $X_1\otimes Z_2$ gate versus $\frac{\gamma}{2}Z_2$ dephasing and $\frac{\xi}{2}Z_1\otimes Z_2$ crosstalk noise. (a) Infidelity for a robust implementation with waveforms satisfying the crosstalk condition given in Eq.~\eqref{eq:crosstalk_condition} and waveforms on the second qubit satisfying the closed-curve condition given in Eq.~\eqref{eq:closed_curve_condition}. Both noise sources are suppressed to first order and thus the infidelity has a quartic dependence in both directions, $\mathcal{I}\propto \mathcal{O}(\gamma^4)  + \mathcal{O}(\xi^4)$. (b) Infidelity for non-robust square pulses of constant amplitude. Neither noise source is suppressed in this case, leading to an infidelity that depends on the noise parameters quadratically in both directions, $\mathcal{I}\propto \mathcal{O}(\gamma^2)  + \mathcal{O}(\xi^2)$. The black dashed line indicates the value of $\mathcal{I} = 10^{-6}$ and serves as a guide to the eye for easier comparison between the two plots. The waveforms are given in Fig.~\ref{fig:Crosstalk_and_dephasing} of the main text. Notice that, for fair comparison, the amplitudes of the non-robust waveforms are chosen such that they are higher than those in the robust implementation.}
\end{figure}

\noindent In this section we derive the crosstalk cancellation condition given in Eq.~\eqref{eq:crosstalk_condition} of the main text. Consider the two-qubit Hamiltonian 
\begin{equation}
    H(t) = H_0(t) +\frac{\xi}{2} Z_1\otimes Z_2 = H_1(t) + H_2(t) + \frac{\xi}{2} Z_1\otimes Z_2,
\end{equation}
where $H_0(t) = H_1(t) + H_2(t)$ is the noise-free Hamiltonian, $H_1(t)$ and $H_2(t)$ are single-qubit control Hamiltonians given by Eq.~\eqref{eq:hamiltonian_single-qubit} with control fields $\{\Omega_i, \Phi_i,\Delta_i\},\, i=1,2$, and $Z_1\otimes Z_2$ is the crosstalk operator with strength $\xi$. Similarly to Sec.~\ref{sec:single-qubit scqc}, we can isolate the effect of the crosstalk term on the evolution by decomposing the evolution operator as $U = U_0U_I$ and then solving the effective Schr\"{o}dinger equation for $U_I$, see Eq.\eqref{eq:schrodinger_eq_noise_operator}. The only difference is that now $U_0 = U_1\otimes U_2$, with $U_i = \mathcal{T}{\rm exp}\{-i \int_0^t d\tau\, H_i(\tau)\}$ being the $i$-th single-qubit evolution operator. To compute the noisy unitary $U_i$, we take its Magnus expansion (controlled by $\xi$) up to first order $U_I(t) \approx e^{-i\Pi_1(t)}$ where  
\begin{equation}
\begin{split}
    \Pi_1(t) &= \frac{\xi}{2}\int_{0}^{t} dt'\, U_0^\dagger(Z_1\otimes Z_2) U_0  = \frac{\xi}{2}\int_0^t dt'\, (U_1^\dagger\otimes U_2^\dagger)(Z_1\otimes Z_2)(U_1\otimes U_2) \\
    &\phantom{=}\\
    &= \frac{\xi}{2}\int_0^t dt'\, (U_1^\dagger Z_1 U_1)\otimes (U_2^\dagger Z_2 U_2) = \frac{\xi}{2} \vec{\sigma}_1^{T}\mathcal{M}\vec{\sigma}_2 \, . 
\end{split}
\label{eq:ZZ_integral}
\end{equation}
where $\vec{\sigma}_i = [X_i, Y_i, Z_i]^T$ is the Pauli vector for each qubit subspace, and $\mathcal{M}$ is a matrix containing time-dependent coefficients. Following the definition of the error curve in Eq.~\eqref{eq:error_curve}, we define the tangent vectors for each qubit to be $\vec{T}_1\cdot \vec{\sigma}_1 = \hat{z}\cdot U_1^\dagger \vec{\sigma}_1 U_1 = U_1^\dagger Z_1U_1$ and $\vec{T}_2(t) = \hat{z}\cdot U_2^\dagger \vec{\sigma}_2 U_2 = U_2^\dagger(t)Z_2U_2(t)$, respectively. Therefore, the geometric representation of the integral in Eq.~\eqref{eq:ZZ_integral} becomes
\begin{equation}
    \mathcal{Q}(t) = \int_0^t dt'\,\vec{T}_1(t')\vec{T}_2(t')\, ,
\end{equation}
where the outer vector product is inferred between the two vectors, $\vec{u}\vec{v}\equiv \vec{u}\otimes\vec{v}^{\,\rm T}$. Our goal is to make $\Pi_1(T_g) = 0$, which is equivalent to making $\mathcal{Q}(T_g) = 0 $. Having two tangent vectors means that we have two 3D space curves, one for each qubit, designed separately, which, in general, do not have the same length and hence the same gate times, i.e., $T_{g1} \neq T_{g2}$. We overcome this issue by normalizing the gate times to $T_{g1} = T_{g2} = T_g = 1$, while the fields for each Hamiltonian are transformed accordingly, for instance $\kappa_i(t) \rightarrow \kappa_i(t)T_{gi}$ and $\tau_i(t) \rightarrow \tau_i(t)T_{gi}$ for $i=1,2$. 
\begin{figure}[b!]
\includegraphics[width=1\linewidth,keepaspectratio]{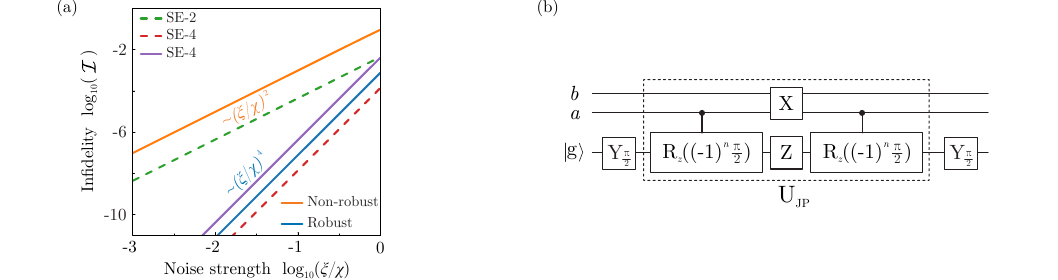}
    \caption{\label{fig:figure_1}\Filippos{(a) Gate infidelity $(\mathcal{I} = 1-\mathcal{F})$ versus crosstalk strength $\xi$ for the $X_1\otimes X_2$ gate. The robust implementation (solid blue) uses square pulses of amplitude $\Omega_1/\chi = \pi$ and $\Omega_2/\chi = 3\pi$, which satisfy the orthogonality condition of Eq.~\eqref{eq:orthogonality_condition} and suppress the crosstalk to leading order. For comparison we show non-robust square pulses (solid orange), spin-echo sequences with two (SE-2, dashed green) and four (SE-4, dashed red) instantaneous $\pi$-pulses, and the same four-pulse sequence with finite-width pulses (SE-4, solid purple). To keep the ratio of field to noise strength comparable across implementations, the non-robust pulses have $\Omega_1/\chi = \Omega_2/\chi = 3\pi$; the finite echo pulses on the second qubit use the same amplitude. (b) Circuit diagram for the joint parity dynamically corrected gate, where $a$ and $b$ are the two physical cavities and $\ket{\textsl{g}}$ is the initial state of the ancilla qubit.}}
\end{figure}
The integral now reads
\begin{equation}
    \mathcal{Q}(1) = \int_0^1dt\, \vec{T}_1(t)\vec{T}_2(t)\,
\label{eq:ZZ_integral_0_1}\, ,
\end{equation}
which is fine in the case where both curves have the same parameterization. However, a space curve $\vec{r}(x)$ with an arbitrary parameterization $x$ can always be re-parameterized to arc-length parameterization using the inverse function of $x^{-1}(t)$. This can be done by integrating the speed, 
\begin{equation}
    \upgamma(x) = \bigg\|\frac{d\vec{r}(x)}{dx}\bigg\|_2= \frac{dt}{dx}\, ,
\end{equation}
of the curve:
\begin{equation}
    t = \int_0^x du\,\upgamma(u)\,.
\end{equation}
The above integration, more often than not, cannot be calculated analytically, but it is straightforward to compute numerically. Therefore, we re-parameterize both curves to arc-length parameterization and continue with Eq.~\eqref{eq:ZZ_integral_0_1}. Notice that re-parameterizing the curve does not affect its shape nor that of its frame vectors. The question now becomes how we can make the quantity $\mathcal{Q}(1)$ vanish. Equation~\eqref{eq:ZZ_integral_0_1} can be written as
\begin{equation}
\label{eq:Q_matrix}
    \mathcal{Q}(1) = \begin{bmatrix}
        \mathcal{M}_{xx} & \mathcal{M}_{xy} & \mathcal{M}_{xz}\\
        \mathcal{M}_{yx} & \mathcal{M}_{yy} & \mathcal{M}_{yz}\\
        \mathcal{M}_{zx} & \mathcal{M}_{zy} & \mathcal{M}_{zz}
    \end{bmatrix}\,, \qquad \qquad \mathcal{M}_{ij}=\int_0^1dt\, T_1^{i}(t)T_2^j(t)\, ,
\end{equation}
where $T^k(t)$ denotes the tangent component along the $k$-axis, $k\in\{x,y,z\}$. So, we can make $\mathcal{Q}(1)$ vanish iff we make all the matrix elements equal to zero, 
\begin{equation}
    \mathcal{M}_{ij}=\int_0^1dt\, T_1^{i}(t)T_2^j(t)=0\, ,\qquad \forall i,j\in \{x,y,z\}\,.
\label{eq:orthogonality_condition}
\end{equation}
This last equation states that by designing two 3D space curves with orthogonal tangent components in the interval $t\in[0,1]$, the quasi-static crosstalk noise is canceled to first order. \\
\indent The crosstalk cancellation condition can be satisfied even for square pulses of constant Rabi strength $\Omega_i$, zero phase field and detuning. These fields result from circular planar arcs of constant curvature, $\kappa_i = \Omega_i$, and tangent vectors $\vec{T}_i(t) = \cos(\kappa_it)\hat{x} + \sin(\kappa_i t)\hat{y}$, with $t\in[0,1]$ (using a rotated coordinate system). The orthogonality condition in Eq.~\eqref{eq:orthogonality_condition} is satisfied simply by choosing $\kappa_1 =\pi$ and $\kappa_2 = 3\pi$. With this choice of curvatures and gate time $T_g = 1$, we implement an $X_1\otimes X_2$ rotation robust to $ZZ$ crosstalk up to first order. The first-order cancellation is shown in Fig.~\ref{fig:figure_1}(a) where the infidelity of the DCG scales with $\xi^4$, while the non-robust gate scales with $\xi^2$. In the case of identity $I_1\otimes I_2$, we can also achieve robustness against dephasing $Z_2$ noise simply by setting $\kappa_1 = 2\pi$ and $\kappa_2 = 4\pi$.\\
\indent In Fig.~\ref{fig:Contour_plot} we plot the average infidelity of an $X_1\otimes Z_2$ gate in the presence of dephasing noise in the second qubit and $ZZ$ crosstalk for a robust implementation that satisfies Eq.~\eqref{eq:dephasing_condition} and Eq.~\eqref{eq:crosstalk_condition}, and a non-robust one that does not satisfy either of them. The Hamiltonian used is similar to Eq.~\eqref{eq:Hamiltonian_step_2} with the only difference being that here the noise terms are taken as $\tilde{h}_n(t)= \frac{\gamma}{2}Z_2 - \frac{\xi}{2}Z_1\otimes Z_2$. The waveforms for both implementations are shown in Fig.~\ref{fig:Crosstalk_and_dephasing} of the main text. Comparing the two plots shown in Fig.~\ref{fig:Contour_plot}, we observe that robust pulses suppress both noise sources to first order thus making the infidelity to not only scale with the fourth power of the noise strengths, $\mathcal{I}\propto \mathcal{O}(\gamma^4)  + \mathcal{O}(\xi^4)$, but also achieve much lower values compared to the non-robust configuration. For instance, non-robust pulses achieve infidelity $\mathcal{I}\leq 10^{-6}$ only for very small noise strengths ($\gamma/\chi\leq 10^{-2.5}$ and $\xi/\chi \leq10^{-2}$), while the robust pulses are below this threshold for almost all the simulated values ($\gamma/\chi \leq 0.5$ and $\xi/\chi \leq 0.5$) (see the dashed line in Fig.~\ref{fig:Contour_plot}).\\
\indent\Filippos{Finally, it is natural to ask whether the $ZZ$ crosstalk considered here could equally well be suppressed by a spin-echo sequence. We address this by introducing instantaneous $\pi$-pulses in the non-robust case where the Rabi rates are chosen constant and equal, i.e. $\Omega_1/\chi = \Omega_2/\chi = 3\pi$, and the target gate is $X_1\otimes X_2$. To refocus the quasi-static $ZZ$, without loss of generality, we choose the second qubit to apply the $\pi$-pulses since $X_2ZZX_2=-ZZ$. For two $\pi$-pulses, at $t_1 = 1/2$ and $t_2 = 1$, the first order in the Magnus expansion (see Eq.~\eqref{eq:ZZ_integral}) reads $\Pi_1(1) = -\frac{\xi}{6\pi}(Y_1Z_2 + Z_1Y_2)$, which is equivalent to suppressing only the diagonal elements of $\mathcal{Q}(1)$. However, the first order in Magnus remains finite and this can be seen also in Fig.~\ref{fig:figure_1}(a) where the infidelity scales with $\propto \xi^2$. The explanation is given by the fact that the integrands of $\mathcal{M}_{yy}$ and $\mathcal{M}_{zz}$ have even symmetry about $t=1/2$, while those of the off-diagonal elements $\mathcal{M}_{yz}$ and $\mathcal{M}_{zy}$ have odd symmetry about the same point. In order to suppress the off-diagonal terms one needs to apply four $\pi$-pulses at times $t_n = n/4$. In this case, the diagonal matrix elements remain suppressed, while the off-diagonal ones are cancelled due to the even symmetry of the sub-intervals about $t=1/4$ and $t=3/4$. Hence, for four pulses we obtain $\Pi_1(1) = 0$ and thus the suppression of the crosstalk to first order. This is also verified through the infidelity shown in Fig.~\ref{fig:figure_1}(a). Of course, what was described above is an idealization, because instantaneous $\pi$-pulses translate to infinite power which is not practical. Finite pulses require finite gate time, therefore the total gate time required for a spin-echo protocol will always be longer than the time required for the geometric approach given in Eq.~\eqref{eq:orthogonality_condition}, even in the case of square pulses. This can be seen by comparing the total angle of rotation between the two approaches, the geometric condition leads to $\vartheta_{\rm SCQC} =3\pi$ while the spin-echo gives $\vartheta_{\rm SE{\text -}4}=\pi + 4\pi = 5\pi$ (the $4\pi$ comes from the four echo pulses). As a result, the spin-echo scheme is slower by a factor of $\vartheta_{\rm SE{\text -}4}/\vartheta_{\rm SCQC} = 5/3$. The situation becomes even worse when the buffer time between driving and echo pulses and the finite rising pulse times must be included. For the sake of comparison, in Fig.~\ref{fig:figure_1}(a) we plot the infidelity for the case of finite echo pulses, which retains the $\propto \xi^4$ scaling but higher compared to our pulses.
}
\section{Dual-rail Hamiltonian truncation \label{sec:dual-rail hamiltonian}}
\noindent In this section, we start from the native Hamiltonian of the system and show that one can truncate it to obtain the effective dual-rail Hamiltonian shown in the main text, following an approach similar to Ref.~\cite{Tsunoda2023}. The native Hamiltonian describes the beam splitter interaction between two bosonic modes along with the dispersive interaction between an ancilla and one of the modes. The Hamiltonian is 
\begin{equation}
    H_{\rm DR}(t) = \frac{g(t)}{2}\left(e^{i\varphi(t)}a^{\dagger}b + e^{-i\varphi(t)}a b^{\dagger}\right) + \delta(t)a^\dagger a - \frac{\chi}{2} a^\dagger a Z_2\, ,
\label{eq:native_hamiltonian}
\end{equation}
where $g(t)$ is the strength and $\varphi(t)$ is the phase of the beam splitter interaction, $\delta(t)$ is an effective mode detuning, $a$ ($a^{\dagger}$) and $b$ ($b^{\dagger}$) are the annihilation (creation) operators acting on the two bosonic modes, $\chi$ is the strength of the dispersive interaction between the ancilla (in the $\textsl{g}$-$f$ manifold) and mode $a$, and $Z_2 \equiv \ket{\textsl{g}}\bra{\textsl{g}} - \ket{f}\bra{f} $ is the Pauli $Z$ operator in the two-level subspace defined by the ground $\ket{\textsl{g}}$ and second excited $\ket{f}$ levels of the transmon ancilla, while the intermediate first excited state $\ket{e}$ is reserved for the error detection of a single ancilla decay event \cite{Rosenblum2018, Reinhold2020}. Considering the dual-rail encoding, where the codewords $\ket{0}_{\rm L}=\ket{10}$ and $\ket{1}_{\rm L}=\ket{01}$ lie in the single-photon subspace of the two-cavity Hilbert space, we can project the full Hamiltonian given in Eq. \eqref{eq:native_hamiltonian} into the single-photon subspace of the two-cavity Hilbert space. Any other state out of this single-photon subspace is considered as leakage out of the codespace and detected by the joint-parity measurement shown in Sec. \ref{sec:three_step parity check}. To do this projection, we use the Schwinger boson representation, where quantum mechanical spin is expressed in terms of two bosonic operators \cite{Schwinger1952} and we define the angular momentum operators
\begin{equation}
    I_1 = a^\dagger a + b^\dagger b\,,\qquad X_1 = a^\dagger b + a b^\dagger\,,\qquad Y_1 =  -i(a^\dagger b - a b^\dagger)\,,\qquad Z_1 = a^\dagger a - b^\dagger b\,.\qquad
\label{eq:angular_momentum_operators}
\end{equation}
and rewrite Eq.\eqref{eq:native_hamiltonian} in the single-photon subspace as:
\begin{equation}
    H_{\rm DR}^{n=1}(t) = \frac{g(t)}{2}(\cos\varphi(t) X_1 - \sin\varphi(t) Y_1) + \frac{\delta(t)}{2}(I_1 + Z_1) - \frac{\chi}{4}(Z_1\otimes Z_2 + Z_2)\, .
\label{eq:dual_rail_hamiltonian_Paulis}
\end{equation}
In the above Hamiltonian, the DR and the ancilla qubits are understood as the first and second qubits, respectively. The Hamiltonian of the main text (see Eq.~\eqref{eq:two_qubit_hamiltonian}) is recovered from Eq.~\eqref{eq:dual_rail_hamiltonian_Paulis} by setting $\varphi = 0$ and $\delta = 0$, 
\begin{equation}
    H_{0}(t) = \frac{g(t)}{2}X_1 - \frac{\chi}{4}Z_1\otimes Z_2 - \frac{\chi}{4}Z_2 + H_2(t)\,,
\label{eq:two_qubit_hamiltonian_DR_and_transmon}
\end{equation}
where we also added the driving on the transmon ancilla through $H_2(t) = \frac{\Omega_2(t)}{2}\left[\cos\Phi_2(t)X_2 + \sin\Phi_{2}(t)Y_2\right] + \frac{\Delta_2(t)}{2}Z_2$ for the sake of completeness. Furthermore, when the beam splitter is not driven ($g(t)=0$), the two cavities are decoupled and by setting the detuning and the phase on the transmon $\Delta_2 =\chi/2 $ and $\Phi_2 = 0$, respectively, the Hamiltonian yields 
\begin{equation}
    H_{0}^{ZZ}(t) = - \frac{\chi}{4}Z_1\otimes Z_2 + \frac{\Omega_2(t)}{2}X_2\,.
\label{eq:ZZ_hamiltoninan}
\end{equation}
Also, by reorganizing Eq.~\eqref{eq:two_qubit_hamiltonian_DR_and_transmon} we obtain the other form of the Hamiltonian used in the main text
\begin{equation}
    H^{XZ}(t) =  H_0^{XZ}(t) - \frac{\chi}{4}Z_1\otimes Z_2\, ,
\label{eq:XZ_hamiltoninan}
\end{equation}
where $H_0^{XZ}(t) = H_1(t) + \tilde{H}_2(t)$ is the noise-free part since $-\frac{\chi}{4}Z_1\otimes Z_2$ is considered noise here, $H_1(t) = \frac{g(t)}{2}X_1$ and $\tilde{H}_2(t) = \frac{\Omega_2(t)}{2}\left[\cos\Phi_2(t)X_2 + \sin\Phi_2(t)Y_2\right] + \frac{\tilde{\Delta}_2(t)}{2}Z_2$  with $\tilde{\Delta}_2(t) = \Delta_2(t) - \frac{\chi}{2}$ are the single-qubit Hamiltonians. Equations ~\eqref{eq:ZZ_hamiltoninan} and \eqref{eq:XZ_hamiltoninan} are used for the joint parity operator and the logical entangling operators. Of course, the Hamiltonians are exposed to transmon-induced dephasing noise $h_n(t) = \frac{\gamma}{2}Z_2$, which we treat with SCQC.
\section{Derivation of the three-step joint parity unitary \label{sec:three_step parity check}}
\noindent As mentioned in the main text, quantum non-demolition (QND) measurement of the two-cavity photon parity is of great importance because, in this way, we can detect leakage errors and convert them to erasure errors when the parity is even, while leaving the qubit unaffected when there is only one photon and the parity is odd. In other words, we can detect that the DR qubit has leaked to the global ground state $\ket{00}$ or any of the two-photon states $\{\ket{20},\ket{11},\ket{02}\}$ and restart the process, or do nothing when the system is in the codespace $\{\ket{01}, \ket{10}\}$. \\
\indent This QND measurement can be executed by leveraging the $ZZ$ interaction and implementing a $Z$ rotation on the ancilla conditional on the two-cavity photon parity. According to Refs.~\cite{Tsunoda2023,Teoh2023}, this operation can be achieved by setting $\delta=0$, $\varphi=0$, and $g = \frac{\sqrt{3}}{2}\chi$ in Eq. \eqref{eq:native_hamiltonian}. By doing this and letting the system evolve for time $T_{\rm JP} = \frac{2\pi}{|\chi|}$ the evolution operator becomes
\begin{equation}
    \widetilde{U}_{\rm JP} = e^{-i\frac{\pi}{2}N}\left[\mathbb{1}\otimes\ket{\textsl{g}}\bra{\textsl{g}} + e^{i\pi N}\otimes \ket{f}\bra{f}\right]\,,
\label{eq:joint_parity_supp}
\end{equation}
where $a^\dagger a + b^\dagger b$ is the total photon number operator. Here, the joint parity gate is executed in a single-shot fashion; however, the design does not take into account the ancilla-induced dephasing noise that leads to an incorrect measurement outcome. Either a valid dual-rail qubit state is unnecessarily erased or a leakage state is not flagged as erasure \cite{Teoh2023}.\\
\indent In order to design a joint parity gate that is robust against the ancilla dephasing noise, we break the operation into three steps where each step dynamically corrects this type of noise up to second order. These steps can be described as (i) letting the ancilla pick up a phase due to the interaction with the photons in mode $a$, (ii) swapping the photons between modes $a$ and $b$, and (iii) letting the ancilla accumulate a phase due to the interaction with the photons that came from mode $b$. The first and third steps are implemented by a $ZZ(\pi/2)$ gate, while the intermediate step is implemented by an $X_1Z_2$ gate, both shown in the main text.
\subsection{\texorpdfstring{$ZZ(\pi/2)$}{TEXT} gate}
We start from Eq.~\eqref{eq:native_hamiltonian} and set $g(t) =0$, $\delta(t)=0$ and project the Hamiltonian onto the subspace $\{\ket{00},\ket{01},\ket{10},\ket{11}\}$ to obtain
\begin{equation}
    H(t) = -\frac{\chi}{2}(\ket{10}\bra{10} + \ket{11}\bra{11})\otimes Z_2\, .
\end{equation}
Next, by adding the transmon control Hamiltonian $H_2(t)$ with $\Phi_2(t) =0 $ and $\Delta_2 = \chi/2$, considering Eq.~\eqref{eq:angular_momentum_operators} for the DR encoding, we obtain the block-diagonal Hamiltonian:
\begin{equation}
\begin{split}
    H_{\rm JP}^{1,3}(t) &= \ket{00}\bra{00} \otimes H_+(t) + \ket{10}\bra{10} \otimes H_-(t) + \ket{01}\bra{01} \otimes H_+(t) + \ket{11}\bra{11} \otimes H_-(t) \\
        &\phantom{=}\\
        &=  \ket{0}\bra{0}_L \otimes H_-(t) + \ket{1}\bra{1}_L \otimes H_+(t) + \ket{00}\bra{00} \otimes H_+(t) + \ket{11}\bra{11} \otimes H_-(t)\, ,
\end{split}
\label{eq:hamiltonian_step_1-3}
\end{equation}
where $\{\ket{0}_{\rm L}, \ket{1}_{\rm L}\}$ are the single-photon states in the DR encoding, $\{\ket{00},\ket{11}\}$ are leakage states out of the DR code space, and 
\begin{equation}
    H_\pm(t) = \frac{\Omega_2(t)}{2} X_2 \pm \frac{\chi}{4}Z_2 + \frac{\gamma}{2}Z_2
\label{eq:Hamiltonian_plus_minus}
\end{equation}
is the noisy single-qubit Hamiltonian with a control field on the transmon. The target evolution operator is:
\begin{equation}
\begin{split}
    V_{\rm JP}^{1,3} &= \ket{0}\bra{0}_L \otimes R_z\left(-\frac{\pi}{2}\right) + \ket{1}\bra{1}_L \otimes R_z\left(\frac{\pi}{2}\right) + \ket{00}\bra{00} \otimes R_z\left(\frac{\pi}{2}\right) + \ket{11}\bra{11} \otimes R_z\left(-\frac{\pi}{2}\right)\\
    &\phantom{=}\\
    &= ZZ\left(\frac{\pi}{2}\right) + \ket{00}\bra{00} \otimes R_z\left(\frac{\pi}{2}\right) + \ket{11}\bra{11} \otimes R_z\left(-\frac{\pi}{2}\right)
\end{split}
\label{eq:U_13}
\end{equation}
where $R_z(\theta) = e^{i\frac{\theta}{2}Z}$ and $ZZ(\theta) = e^{-i\frac{\theta}{2} Z_1\otimes Z_2}$. Given that the dispersive coupling $\chi$ is constant and serves as the effective detuning for the transmon, the only degree of freedom left is the transmon Rabi rate. So, we need a control field $\Omega_2(t)$ such that the transmon undergoes the conditional rotations $R_z(\pm \frac{\pi}{2})$ while dynamically suppressing the dephasing noise introduced by $\gamma$ to first order. As shown in the main text, in SCQC this translates to a closed curve of constant torsion where the relative angle of the frame vectors is equal to $\theta = \pi/2$ around the axis defined by the tangent vector.
\subsection{\texorpdfstring{$X_1\otimes Z_2$}{TEXT} gate}
\indent For the $X_1\otimes Z_2$ unitary, we start from Eq.~\eqref{eq:native_hamiltonian}, we set $\delta(t)=0$ and restrict $g(t)$ to waveforms that only implement a one-photon swap between the cavities, like the Gaussian pulse proposed in the main text. Then, we project the Hamiltonian onto the subspace $\{\ket{00},\ket{01},\ket{10},\ket{11}\}$ and obtain 
\begin{equation}
    H_{\rm JP}^{2}(t) =  \frac{g(t)}{2}X_1  + \tilde{H}_2(t) + \tilde{h}_n(t)\, ,
\label{eq:Hamiltonian_step_2}
\end{equation}
where $\tilde{H}_2(t)$ is the ancilla control Hamiltonian given right below Eq.~\eqref{eq:XZ_hamiltoninan}, and $\tilde{h}_n(t)= \frac{\gamma}{2}Z_2 - \frac{\xi}{4}Z_1\otimes Z_2$ contains the dephasing and crosstalk noise terms. We replaced $\chi$ with $\xi(t)$ to account for the general case of unknown and possibly stochastic noise strength. By imposing this restriction to $g(t)$ we ensure that the population remains within the projected space. In the main text, we choose a Gaussian pulse for $g(t)$ that satisfies the restriction and leads to an $X_1$ gate, while for the transmon ancilla we employ SCQC to design a closed curve that suppresses crosstalk noise by satisfying the orthogonality condition given in Eq.~\eqref{eq:orthogonality_condition}, while implementing a $Z_2$ gate. Therefore, the \Filippos{truncated} evolution operator is
\begin{equation}
    V_{\rm JP}^{2} =  (\ket{00}\bra{00} - \ket{11}\bra{11} -i X_1)\otimes Z_2\, ,
\label{eq:U_2}
\end{equation}
where the dephasing and crosstalk terms are suppressed to first order and thus not included.\\
\indent\Filippos{The $X_1$ factor in Eq.~\eqref{eq:U_2} originates from the full beam-splitter operation between the two cavities. For a BS pulse of area $\int g(t)\,\mathrm{d}t = \pi$, the cavity evolution operator in the basis $\{\ket{00},\ket{01},\ket{10},\ket{11},\ket{02},\ket{20}\}$ is
\begin{equation}
    U_{\rm BS} = \exp\left[-i\frac{\pi}{2}\left(a^\dagger b + a b^\dagger\right)\right] = e^{-i\frac{\pi}{2}N}\,\mathcal{S}_{ab} =
    \begin{pmatrix}
        1 & 0 & 0 & 0 & 0 & 0 \\
        0 & 0 & -i & 0 & 0 & 0 \\
        0 & -i & 0 & 0 & 0 & 0 \\
        0 & 0 & 0 & -1 & 0 & 0 \\
        0 & 0 & 0 & 0 & 0 & -1 \\
        0 & 0 & 0 & 0 & -1 & 0
    \end{pmatrix}\, ,
\label{eq:BS_swap}
\end{equation}
where $\mathcal{S}_{ab}$ swaps the photon numbers of the two cavities, $\mathcal{S}_{ab}\ket{nm} = \ket{mn}$, and $N = a^\dagger a + b^\dagger b$.
The block structure shows that the truncated subspace $\{\ket{00},\ket{01},\ket{10},\ket{11}\}$ is mapped onto itself, where $U_{\rm BS}$ reduces to $\ket{00}\bra{00} - \ket{11}\bra{11} - iX_1$, with $X_1$ acting on the single-excitation DR subspace $\{\ket{01},\ket{10}\}$. Note that during the operation, the state $\ket{11}$ is not confined to the truncated subspace, but evolves through the two-photon states $\ket{02}$ and $\ket{20}$ before returning to $-\ket{11}$ at the end of the pulse. All simulations in the main text and the Supplemental Information include the cavity states up to two photons, $\left\{\ket{11},\ket{02},\ket{20}\right\}$, and therefore account for the population of $\ket{02}$ and $\ket{20}$ during the BS operation. The projection onto the truncated subspace, $\left\{\ket{00},\ket{01},\ket{10},\ket{11}\right\}$, is applied only at the end of the complete joint parity and logical entangling operations.}
\subsection{Joint parity gate}
Finally, we show how the robust unitaries derived above implement the joint parity unitary when applied sequentially:
\begin{equation}
\begin{split}
    U_{\rm JP} &= V_{\rm JP}^{3}V_{\rm JP}^{2}V_{\rm JP}^{1} \\
               &\phantom{=}\\
               &= V_{\rm JP}^{3}\left((-iX_1 \otimes Z_2) ZZ\left(\frac{\pi}{2}\right) + \ket{00}\bra{00} \otimes Z_2R_z\left(\frac{\pi}{2}\right) - \ket{11}\bra{11} \otimes Z_2R_z\left(-\frac{\pi}{2}\right)\right) \\
               &\phantom{=}\\
               &= -iZZ\left(\frac{\pi}{2}\right) (X_1\otimes Z_2) ZZ\left(\frac{\pi}{2}\right) + \ket{00}\bra{00} \otimes R_z\left(\frac{\pi}{2}\right)Z_2R_z\left(\frac{\pi}{2}\right) - \ket{11}\bra{11} \otimes R_z\left(-\frac{\pi}{2}\right)Z_2R_z\left(-\frac{\pi}{2}\right)\\
               &\phantom{=}\\
               &= i\left(\ket{00}\bra{00} + \ket{11}\bra{11}\right)\otimes I_2 - iX_1\otimes Z_2 = i\left[\left(P_{00}+P_{11}\right)\otimes\mathbb{1} - X_1\otimes Z_2\right]\, ,
\end{split}
\label{eq:Joint_parity_final}
\end{equation}
where we used $(X_1\otimes I_2) ZZ(\theta)(X_1\otimes I_2) = ZZ(-\theta)$ and   $ZZ(\theta)ZZ(-\theta) = I_1\otimes I_2$. \Filippos{Of course, the protocol can be extended to higher photon numbers, with the ancilla undergoing an identity $I_2$ operation for even parity and a $Z_2$ rotation for odd parity, although robust operation requires redesigning the pulses to account for the additional photon-number states.} This step-wise implementation leads to a logical $X$ rotation on the DR qubit, which can be easily tracked in software or undone at the end of the erasure check. The performance of the robust parity check operation against ancilla-dephasing is shown in Fig.~\ref{fig:Parity_and_ZZgate}(b), revealing the noise suppression to first order until a plateau is reached when the $ZZ$ crosstalk noise becomes the dominant noise source in the system. Furthermore, the general protocol with higher photon number in the cavities and the ancilla state preparation can be seen in Fig.~\ref{fig:figure_1}(b), where a SWAP operation is implemented between the cavities' modes. \\
\subsection{Second-order cancellation \label{sec:second_order_cancellation}}
Here, we show that the $X_1\otimes Z_2$ gate  implemented in step (ii) of our robust joint parity check protocol cancels the ancilla-induced dephasing noise to second order when the system is in the codespace $\{\ket{01}, \ket{10}\}$. In the codespace, the Hamiltonian for steps (i) and (iii) is given by Eq.~\eqref{eq:hamiltonian_step_1-3}:
\begin{equation}
        H_{\rm JP}^{\rm 1,3}(t) =  \ket{0}\bra{0}_L \otimes \big[H_{0,-}(t) + \frac{\gamma}{2}Z_2\big] + \ket{1}\bra{1}_L \otimes \big[H_{0,+}(t)+ \frac{\gamma}{2}Z_2\big] \, ,
\label{eq:Hamiltonian_pm_for_XZ_section}
\end{equation}
where $H_{0,\pm}(t) = \frac{1}{2}[\Omega_2(t)X_2 \pm \frac{\chi}{2} Z_2]$ are the noise-free Hamiltonians for each subspace. The total evolution operator is written as $V_{\rm JP}^{1,3} = V^{1,3}_{0} V^{1,3}_{I}$, where the noiseless evolution operator is given as
\begin{equation}
    V^{1,3}_0(t) = \ket{0}\bra{0}_L \otimes \mathcal{T}e^{ -i\int_0^tdt'H_{0,-}(t')} + \ket{1}\bra{1}_L \otimes \mathcal{T}e^{ -i\int_0^tdt'H_{0,+}(t')} = \ket{0}\bra{0}_L \otimes U_{0,-}(t) + \ket{1}\bra{1}_L \otimes U_{0,+}(t)\, ,
\end{equation} 
while the noise evolution operator is given by $V^{1,3}_I(t) = \mathcal{T}{\rm exp}\{-i\int_0^tdt'\,\tilde{h}_n(t')\}$, where 
\begin{equation}
\begin{split}
     \tilde{h}_n(t) = \big[V^{1,3}_{0}(t)\big]^\dagger\frac{\gamma}{2}Z_2\big[V^{1,3}_{0}(t)\big] &=  \frac{\gamma}{2} \Big[\ket{0}\bra{0}_L \otimes U_{0,-}^\dagger(t)Z_2 U_{0,-}(t) + \ket{1}\bra{1}_L \otimes U_{0,+}^\dagger(t)Z_2 U_{0,+}(t) \Big]\\
     & \phantom{=}\\
     & = \frac{\gamma}{2} \Big[ -\ket{0}\bra{0}_L \otimes X_2 U_{0,+}^\dagger(t)Z_2 U_{0,+}(t)X_2 + \ket{1}\bra{1}_L \otimes U_{0,+}^\dagger(t)Z_2 U_{0,+}(t)\Big]\\
     & \phantom{=}\\
     & = \frac{\gamma}{2} \Big[ - \ket{0}\bra{0}_L \otimes X_2 h_I(t) X_2 + \ket{1}\bra{1}_L \otimes h_I(t)\Big]\,,
\end{split}
\end{equation}
where we used the fact that $H_{0,-}(t) = X_2 H_{0,+}(t)X_2$, and also defined $h_{I}(t) = U^\dagger_{0,+}(t)Z_2U_{0,+}(t)$. Up to second order in the Magnus expansion \cite{Blanes2009}, the noise evolution operator reads
\begin{equation}
    V^{1,3}_I(t) \approx e^{-i\Pi_{1}(t) - i\Pi_{2}(t) }\, ,
\label{eq:Magnus_second_order}
\end{equation}
with
\begin{subequations}
\begin{equation}
    \Pi_1(t)= \int_0^t dt_1\, \tilde{h}_n(t_1) = \frac{\gamma}{2}\bigg\{-\ket{0}\bra{0}_{\rm L}\otimes X_2\int_0^t dt_1 h_I(t_1) X_2 + \ket{1}\bra{1}_{\rm L}\otimes \int_0^t dt_1 h_I(t_1)  \bigg\}\, ,\phantom{mmmmmmmm}
\label{eq:magnus_first_term}
\end{equation}
\begin{equation}
\begin{split}
    \Pi_2(t) &= -\frac{i}{2}\int_{0}^{t}\int_{0}^{t_1}dt_1dt_2\, [\tilde{h}_n(t_1), \tilde{h}_n(t_2)] \\
    & \phantom{=}\\
    &= -i\frac{(\gamma/2)^2}{2} \bigg\{\ket{0}\bra{0}_{\rm L}\otimes X_2\int_{0}^{t}\int_{0}^{t_1}dt_1 dt_2[h_I(t_1), h_I(t_2)] X_2 + \ket{1}\bra{1}_{\rm L}\otimes \int_{0}^{t}\int_{0}^{t_1}dt_1 dt_2[h_I(t_1), h_I(t_2)] \bigg\}\,.
\end{split}
\label{eq:magnus_second_term}
\end{equation}
\end{subequations}
To suppress dephasing noise up to second order (i.e., $V^{1,3}_I(T_g) \approx I$) we need to make $\Pi_1(T_g) = 0 = \Pi_2(T_g)$. In the SCQC framework, closed curves guarantee the condition $\Pi_1(T_g) =0$, while closed curves with vanishing-area planar projections satisfy $\Pi_2(T_g) = 0$ \cite{Zeng2019}. In our case, $\Pi_1(T_g)=0$ is already satisfied due to the closed curve presented in the main text and shown in Fig.~\ref{fig:Pulse_and_robustness}(a). Recall that the tangent vector is defined as $\vec{T}(t)\cdot \vec{\sigma}_2 =U^\dagger_{0,+}(t)Z_2U_{0,+}(t)$, thus
\begin{equation}
    \int_{0}^{t}\int_{0}^{t_1}dt_1 dt_2[h_I(t_1), h_I(t_2)] = -2i\int_0^{t}dt_1 [\vec{r}(t_1)\times \vec{T}(t_1)] \cdot \vec{\sigma}_2= -2i\vec{A}(t) \cdot \vec{\sigma}_2  \, ,
\end{equation}
where $\vec{A}(t) = [A_x(t),A_y(t),A_z(t)]$, with the components being proportional to the areas enclosed by the closed curve projected onto the $yz$, $zx$, and $xy$ planes, respectively. Now we can rewrite Eq.~\eqref{eq:magnus_second_term} in a more compact form
\begin{equation}
    \Pi_2(t) = -\frac{\gamma^2}{4}\Big\{ \ket{0}\bra{0}_{\rm L}\otimes X_2 \hat{A}(t)X_2 + \ket{1}\bra{1}_{\rm L}\otimes \hat{A}(t) \Big\}\,,
\end{equation}
where $\hat{A}(t) \equiv \vec{A}(t)\cdot \vec{\sigma}_2$, and rewrite the noise evolution operator as $V^{1,3}_I(t) \approx e^{-i\Pi_2(t)}$.\\
\indent Next, we write the joint parity operation [see Eq.~\eqref{eq:U_13}] within the codespace while assuming an ideal $X_1\otimes Z_2$ gate during step (ii) of our protocol
\begin{equation}
\begin{split}
    U_{\rm JP} = V_{\rm JP}^{3}(T_g)X_1\otimes Z_2 V_{\rm JP}^{1}(T_g) &= V_{0}^{3}(T_g) V^{3}_I(T_g) X_1\otimes Z_2 V^{3}_0(T_g)V^{3}_I(T_g) \\
    &\phantom{=}\\
    &=  (V_{0}^{3}X_1\otimes Z_2 V^{3}_0) \big[(X_1\otimes Z_2 V^{3}_0)^\dagger V^{3}_I (X_1\otimes Z_2 V_0^{3})\big]V^{3}_I\, ,
\end{split}
\label{eq:second_order_cancellation_ideal_XZ}
\end{equation}
where we dropped the gate time, $T_g$, for simplicity and used the fact that $V_{\rm JP}^3 =V^{3}_0V^3_I= V^{1}_0V^1_I =V_{\rm JP}^1$. Also, the first parenthesis is the noise-free part, while the rest is noisy and can be written as
\begin{equation}
\begin{split}
    (X_1\otimes Z_2 V^{3}_0)^\dagger V^{3}_I (X_1\otimes Z_2 V^{3}_0) V^{3}_I &= e^{i\frac{\gamma^2}{4}V^{3\dagger}_0X_1\otimes Z_2\big[ \ket{0}\bra{0}_{\rm L}\otimes X_2 \hat{A}(T_g)X_2 + \ket{1}\bra{1}_{\rm L}\otimes \hat{A}(T_g) \big]X_1\otimes Z_2V^{3}_0}V^3_I\\
    &\phantom{=}\\
    &= e^{i\frac{\gamma^2}{4}\big[ \ket{1}\bra{1}_{\rm L}\otimes e^{-i\frac{\pi}{4}Z_2} Z_2 X_2 \hat{A}(T_g)X_2Z_2 e^{i\frac{\pi}{4}Z_2} + \ket{0}\bra{0}_{\rm L}\otimes e^{i\frac{\pi}{4}Z_2} Z_2  \hat{A}(T_g)Z_2 e^{-i\frac{\pi}{4}Z_2}\big]}e^{-i\Pi_2(T_g)}\\
    &\phantom{=}\\
    &= e^{-i \left\{\Lambda + \Pi_2(T_g) - \frac{i}{2}[\Lambda, \Pi_2(T_g)] - \frac{1}{12}[\Lambda,[\Lambda, \Pi_2(T_g)]] - \frac{1}{12}[\Pi_2(T_g), [\Pi_2(T_g),\Lambda]] + \dots \right\} }\\
    &\phantom{=}\\
    &\approx e^{-i \left\{\Lambda + \Pi_2(T_g)\right\}-i\mathcal{O}(\gamma^4)}\, ,
\end{split}
\label{eq:noise_term_XZ}
\end{equation}
where we used that $V^3_0 = ZZ(\pi/2) = e^{-i\frac{\pi}{4}Z1\otimes Z_2}$ [see Eq.~\eqref{eq:U_13}], and approximated the product of the exponentials to the first order, ie $e^{-i \Lambda} e^{-i\Pi_2(T_g)} \approx e^{-i[\Lambda + \Pi_2(T_g)]}$ with $\Lambda$ being the exponent of the first term. The noise term described in Eq.~\eqref{eq:noise_term_XZ} becomes the identity when the exponent is zero, hence for suppressing ancilla dephasing noise to second order it remains only to show that $ \Lambda + \Pi_2(T_g) = 0$: 
\begin{equation}
\begin{split}
     \Lambda + \Pi_2(T_g) &= -\frac{\gamma^2}{4} \Big[ \ket{1} \bra{1}_{\rm L}\otimes e^{-i\frac{\pi}{4}Z_2} Z_2 X_2 \hat{A}(T_g)X_2Z_2 e^{i\frac{\pi}{4}Z_2} + \ket{0}\bra{0}_{\rm L}\otimes e^{i\frac{\pi}{4}Z_2} Z_2  \hat{A}(T_g)Z_2 e^{-i\frac{\pi}{4}Z_2} + \Pi_2(T_g)\Big] \\
    &\phantom{=}\\
    &=-\frac{\gamma^2}{4}\Big[  \ket{1}\bra{1}_{\rm L}\otimes \mathcal{S}(T_g) +  \ket{0} \bra{0}_{\rm L}\otimes X_2\mathcal{S}(T_g)X_2\Big]\,,
\end{split}
\end{equation}
with 
\begin{equation}
    \mathcal{S}(T_g) = e^{-i\frac{\pi}{4}Z_2}Z_2X_2\hat{A}(T_g)X_2Z_2e^{i\frac{\pi}{4}Z_2} + \hat{A}(T_g) = [-A_y + A_x, -A_x + A_y, 0]\cdot \vec{\sigma}_2 \, .
\label{eq:second_order_suppression_symmetry}
\end{equation}
In our case, $\mathcal{S}(T_g) = 0$ due to the symmetry of the designed closed curve (see Fig.~\ref{fig:Pulse_and_robustness}) that leads to $A_x = A_y$. The effect of the second-order cancellation in the dephasing strength is shown in Fig.~\ref{fig:False_negative_main}, where the false positive probability scales like $\propto \gamma^6$ compared to the $\propto \gamma^4$ slope in the case of false negative probability. Of course, this effect affects only the codespace and thus cannot be observed through the average fidelity of the whole joint parity operator. \Filippos{ Although Eq.~\eqref{eq:second_order_suppression_symmetry} was derived assuming an ideal intermediate step $X_1\otimes Z_2$ in Eq.~\eqref{eq:second_order_cancellation_ideal_XZ}, the analysis and the resulting second-order suppression also hold for the finite-duration robust pulses used in the main text [see Fig.~\ref{fig:Crosstalk_and_dephasing}]. This is because the step (ii) pulses are themselves robust to dephasing to first order, and their duration is much shorter than that of steps (i) and (iii). The situation is analogous to spin-echo sequences, where pulses of finite width still suppress noise as long as they are short compared with the free evolution. 
}
\Filippos{
\section{Noise suppression with spin-echo pulses \label{sec:spin_echo}}
\noindent In this section, we provide an alternative approach to suppressing quasi-static dephasing and crosstalk noise using spin-echo (SE) pulses. Furthermore, we explore their limitations, and compare and contrast their performance with the SCQC approach and the waveforms provided in the main text. For convenience in the analysis and fair comparison between the two approaches, we use the same three-part composite sequence proposed in the main text with the only difference being that we now attempt to implement each step with SE pulses with the aim of suppressing the noise while implementing the target gates shown in Fig.~\ref{fig:circuits}(b). We show that this can be done for step (ii) of the sequence, but not for steps (i) and (iii). Thus, we will only replace our SCQC pulses with SE pulses in step (ii) and compare the result to our fully SCQC-based sequence.\\
\subsection{Spin-echo and \texorpdfstring{$ZZ(\pi/2)$}{TEXT}}
\noindent The Hamiltonian that implements the $ZZ(\pi/2)$ gate in steps (i) and (iii) is given in Eq.~\eqref{eq:Hamiltonian_plus_minus}. From Eq.~\eqref{eq:Hamiltonian_plus_minus} it becomes obvious that the application of instantaneous echo pulses along the $x$ or $y$ axis can suppress the dephasing noise term $(\gamma/2) Z_2$; however, the same pulses will refocus the conditional interaction given by $\pm (\chi/4) Z_2$. Thus, we can cancel the noise, but we cannot implement the target entangling gate $ZZ(\pi/2)$ while doing so. This observation shows that instantaneous SE pulses cannot distinguish between the desired interaction and the noise terms, a limitation intrinsic to the echo approach. In contrast, the discrimination between the desired dynamics and the noise sources to be suppressed is one of the core features of the SCQC framework. Recall that we take advantage of this property in our design where we are using the dispersive interaction as a resource in steps (i) and (iii) while suppressing it in step (ii). 
\subsection{Spin-echo and \texorpdfstring{$X_1\otimes Z_2$}{TEXT}}
\label{sec:spin_echo_X1Z2}
Now we turn our focus to step (ii) (recall Fig.~\ref{fig:circuits}(b)). Our goal is to implement the $X_1\otimes Z_2$ gate while suppressing transmon dephasing and crosstalk noise through SE pulses. The Hamiltonian we use for this is similar to Eq.~\eqref{eq:Hamiltonian_step_2}, namely
\begin{equation}
    H_{\rm JP}^{2,\rm SE}(t) = \frac{g(t)}{2}X_1 + \frac{\Omega_{\rm SE}(t)}{2}X_2 + \tilde{h}_n(t)\, ,
\end{equation}
where $g(t)$ and $\Omega_{\rm SE}(t)$ are the driving fields for the beamsplitter and the transmon, respectively, while $\tilde{h}_n(t)= \frac{\gamma}{2}Z_2 - \frac{\xi}{4}Z_1\otimes Z_2$ describes the dephasing and crosstalk noise terms to be suppressed. Without loss of generality, we take the gate time to be $T_g = 1$ and define the accumulated angles 
\begin{equation}
    \vartheta(t) = \int_{0}^{t}g(\tau)d\tau \, , \qquad\qquad \varphi(t)= \int_0^t \Omega_{\rm SE}(\tau)d\tau\, ,
\label{eq:accumulated_angles_1}
\end{equation}
so the noiseless evolution operator is given $U_0(t) = e^{-i\vartheta(t)/2X_1}\otimes e^{-i\varphi(t)/2X_2}$. By setting $\vartheta(1) = \pi$ and $\varphi(1) = 2k\pi$ with $k\in \mathbb{Z}$, and using a virtual $Z_2$ rotation, we can implement the target gate $U_0(1) = X_1\otimes Z_2$ up to a global phase. To that end, we drive the BS with constant strength, and we apply signed $\pi$-pulses on the transmon such that 
\begin{equation}
    \vartheta(t) = \pi t \, , \qquad\qquad \varphi(1) = \pi\sum_{j=1}^{N}s_j\, ,
\label{eq:accumulated_angles_2}
\end{equation}
where $N$ is the total number of pulses, and $s_j$ is the sign of the $j$-th pulse on the ancilla. For practical purposes we choose to have the SE pulses on the transmon because we can drive it more strongly than the BS and thus fit more echo pulses of finite width into the same fixed time window.\\
\indent To get the total evolution operator $U = U_0U_I$ we follow the analysis described in Eqs.~\eqref{eq:schrodinger_eq_noise_operator}-\eqref{eq:dephasing_condition}. Here, $U_I$ is generated by $\tilde{H}_n =U_0^{\dagger}(t)\tilde{h}_n(t)U_0(t)$, while the first order in the Magnus expansion at the end of the evolution is
\begin{equation}
    \Pi_1(t=1)= \int_0^1 \tilde{H}_n(t)dt = \frac{\gamma}{2}\left(\mathcal{I}_Y Y_2 + \mathcal{I}_Z Z_2\right) - \frac{\xi}{4}\left(\mathcal{I}_{ZY}Z_1\otimes Y_2 + \mathcal{I}_{YZ}Y_1\otimes Z_2 + \mathcal{I}_{ZZ}Z_1\otimes Z_2 +\mathcal{I}_{YY}Y_1\otimes Y_2 \right)\, ,
\label{eq:magnus_term_SE}
\end{equation}
with the first two integrals quantifying the effect of dephasing noise:
\begin{equation}
    \mathcal{I}_{Y}=\int_0^1\sin(\varphi)dt\, , \qquad\qquad \mathcal{I}_{Z}=\int_0^1\cos(\varphi)dt\, ,
\label{eq:dephahing_integrals_SE}
\end{equation}
and the remaining four of them quantifying crosstalk errors:
\begin{equation}
    \mathcal{I}_{ZY}=\int_0^1\cos(\vartheta)\sin(\varphi)dt\, ,\quad \mathcal{I}_{YZ}=\int_0^1\sin(\vartheta)\cos(\varphi)dt\, ,\quad \mathcal{I}_{YY}=\int_0^1\sin(\vartheta)\sin(\varphi)dt\, ,\quad \mathcal{I}_{ZZ}=\int_0^1\cos(\vartheta)\cos(\varphi)dt\, .
\label{eq:crosstalk_integrals_SE}
\end{equation}
To lowest order, the average gate infidelity \cite{Nielsen2002, Pedersen2007} is
\begin{equation}
    1-\mathcal{F} = \frac{d^2-|{\rm tr}(U_I)|^2}{d(d+1)} = \frac{\rm{tr}(\Pi_1^2)}{d+1} = \gamma^2\frac{1}{5}\left(\mathcal{I}_{Z}^2 +\mathcal{I}_{Y}^2\right) +  \xi^2\frac{1}{20}\left(\mathcal{I}_{ZY}^2+\mathcal{I}_{YZ}^2 + \mathcal{I}_{ZZ}^2 + \mathcal{I}_{YY}^2\right)\, ,
\label{eq:infidelity_SE}
\end{equation}
where we set $d=4$ as we are now focusing on the Hilbert space defined by the DR code space and the ancillary transmon. Equation~\eqref{eq:infidelity_SE} shows that first-order suppression of dephasing requires the integrals in Eq.~\eqref{eq:dephahing_integrals_SE} to vanish, while first-order suppression of crosstalk requires the integrals in Eq.~\eqref{eq:crosstalk_integrals_SE} to vanish.\\
\indent Given the constant pulse on the BS and the requirement of an even number of echo pulses on the transmon, it is convenient to assume a sequence that is symmetric about $t=1/2$. Assuming that the $j$-th echo pulse is applied at time $t_j$ with sign $s_j$, the symmetry condition imposes a palindromic pattern of times $t_{N+1-j} = 1-t_j$ and signs $s_{N+1-j} = s_j$. With these choices of $\vartheta(t)$ and $\varphi(t)$, $\mathcal{I}_{Y}$, $\mathcal{I}_{ZZ}$, and $\mathcal{I}_{YY}$ vanish identically, since their integrands are odd about $t=1/2$. We are therefore left with finding times $t_j$ and signs $s_j$ that make the remaining integrals, $\mathcal{I}_{Z}$, $\mathcal{I}_{ZY}$, and $\mathcal{I}_{YZ}$, vanish. As we show later, for ideal pulses of infinite amplitude $\mathcal{I}_{ZY}$ vanishes, since $\sin\varphi(t) = 0$ ($\varphi(t) = k\pi$, $k\in\mathbb{Z}$). However, for pulses of finite amplitude and width this no longer holds identically, and $\mathcal{I}_{ZY}$ must be cancelled explicitly.\\
\indent In the case of instantaneous pulses, $\sin(\varphi) = 0$ and $\cos(\varphi) = \pm 1$. Note that $\cos(\varphi)$ flips sign at every pulse, regardless of the pulse sign $s_j$, since $\cos(\varphi \pm \pi) = -\cos(\varphi)$. Moreover, $\mathcal{I}_{ZY}$ vanishes in this limit, so only $\mathcal{I}_{Z}$ and $\mathcal{I}_{YZ}$ remain. We define
\begin{equation}
    G(t) = \int_0^t \sin\big(\vartheta(t')\big)\,\mathrm{d}t' = \frac{1-\cos(\pi t)}{\pi}\, ,
\label{eq:G_1}
\end{equation}
which satisfies
\begin{equation}
    G(1-t) = G(1) - G(t) = 2G\left(\tfrac{1}{2}\right) - G(t)\, ,
\label{eq:G_2}
\end{equation}
with $G_(1) = 2G(1/2) = 2/\pi$. We now solve $\mathcal{I}_{Z} = \mathcal{I}_{YZ} = 0$ for different numbers of pulses.\\
For $N=2$, the pulses are applied at $t_1$ and $1-t_1$, the time intervals between the pulses are $t_1$, $1-2t_1$, $t_1$, and $\cos(\varphi) = +1,-1,+1$ in these intervals. Using these relations, we obtain
\begin{equation}
    \mathcal{I}_{Z}^{(2)} = 4t_1 - 1\, , \qquad\qquad \mathcal{I}_{YZ}^{(2)} = 4G(t_1) - 2G\left(\tfrac{1}{2}\right)\, .
\label{eq:N_2_conditions}
\end{equation}
For $N=4$, the pulses are applied at $t_1$, $t_2$, $1-t_2$, $1-t_1$, the time intervals are $t_1$, $t_2-t_1$, $1-2t_2$, $t_2-t_1$, $t_1$, and $\cos(\varphi) = +1,-1,+1,-1,+1$. The integrals are
\begin{equation}
    \mathcal{I}_{Z}^{(4)} = 4(t_1-t_2) + 1\, , \qquad\qquad \mathcal{I}_{YZ}^{(4)} = 4\left[G(t_1)-G(t_2)\right] + 2G\left(\tfrac{1}{2}\right)\, .
\label{eq:N_4_conditions}
\end{equation}
For $N=6$, following the same analysis, we obtain
\begin{equation}
    \mathcal{I}_{Z}^{(6)} = 4(t_1-t_2+t_3) - 1\, , \qquad\qquad \mathcal{I}_{YZ}^{(6)} = 4\left[G(t_1)-G(t_2)+G(t_3)\right] - 2G\left(\tfrac{1}{2}\right)\, .
\label{eq:N_6_conditions}
\end{equation}
By solving Eq.~\eqref{eq:N_2_conditions} we obtain 
\begin{equation}
    \mathcal{I}_Z^{(2)} = 0 \Rightarrow t_1 = \frac{1}{4}\,, \qquad\qquad \mathcal{I}_{YZ}^{(2)} =0 \Rightarrow t_1 = \frac{1}{3}\, ,
\label{eq:N_2_solution}
\end{equation}
revealing that two echo pulses cannot suppress both noise sources simultaneously. Next, for $N=4$, we impose the vanishing conditions for Eq.~\eqref{eq:N_4_conditions} and obtain the times
\begin{equation}
    t_1 = \frac{1}{\pi}\arcsin\left[\frac{1}{4\sin\left(\pi/8\right)}\right] - \frac{1}{8}\,, \qquad\qquad t_2 = t_1 + \frac{1}{4}\,,
\label{eq:N_4_solutions}
\end{equation}
i.e., $t_1 \approx 0.102$ and $t_2 \approx 0.352$. The second solution of the sine equation, $\pi t_1 + \pi/8 = \pi - \alpha$ with $\alpha = \arcsin\{[4\sin(\pi/8)]^{-1}\}$, gives $t_1' = 1 - t_2$ and $t_2' = 1 - t_1$. This corresponds to the same set of pulse times, $\{t_1, t_2, 1-t_2, 1-t_1\}$, with the pulses in the two halves of the sequence relabeled, and therefore does not constitute a distinct solution. Choosing the principal branch ensures the ordering $t_1 < t_2 < 1/2$. This result shows that, with the imposed symmetry, four infinitely strong pulses applied at the times given by Eq.~\eqref{eq:N_4_solutions} make all coefficients in Eqs.~\eqref{eq:dephahing_integrals_SE}--\eqref{eq:crosstalk_integrals_SE} vanish. Solving the suppression conditions for $N=6$ is therefore unnecessary at this stage, since the system of two equations with three free times is underdetermined. However, this additional degree of freedom will prove useful when we consider pulses of finite amplitude.\\
\indent From now on, we consider pulses of finite amplitude. Since each of the $j-1$ preceding pulses rotates $\varphi$ by $\pm\pi$, the phase at the beginning of the $j$-th pulse is $\varphi = (j-1)\pi$ modulo $2\pi$. We take the $j$-th pulse to be a square pulse of width $w$ and amplitude $\Omega_{\rm SE} = s_j\pi/w$, centered at $t_j$, and define the local time $\tau = t - t_j \in [-w/2, w/2]$. The phase during the pulse is then
\begin{equation}
    \varphi_j(\tau) = (j-1)\pi + s_j\pi\left(\frac{\tau}{w} + \frac{1}{2}\right)\, ,
\label{eq:phase_finite_pulse}
\end{equation}
which gives $\sin(\varphi) = (-1)^{j-1}s_j\cos(\pi\tau/w)$ and $\cos(\varphi) = (-1)^{j}\sin(\pi\tau/w)$. Since $\sin(\varphi)$ no longer vanishes during the pulses, $\mathcal{I}_{ZY}$ in Eq.~\eqref{eq:crosstalk_integrals_SE} is not identically zero for pulses of finite width. Also $\cos(\varphi)$ is odd in $\tau$ within each pulse, its integral over every pulse window vanishes, exactly as in the case of instantaneous pulses, and therefore $\mathcal{I}_{Z}^{(N)}$ is still given by Eqs.~\eqref{eq:N_2_conditions}--\eqref{eq:N_6_conditions}. Note that $\sin(\varphi)$ vanishes outside the pulse windows, thus only the pulses contribute to $\mathcal{I}_{ZY}$ in Eq.~\eqref{eq:crosstalk_integrals_SE}, and we obtain
\begin{equation}
    \mathcal{I}_{ZY} = \sum_j \int_{-w/2}^{w/2}(-1)^{j-1}s_j\cos\left(\frac{\pi\tau}{w}\right)\cos\left[\pi(t_j + \tau)\right]\mathrm{d}\tau = \sum_j y_jc_j\int_{-w/2}^{w/2}\cos\left(\frac{\pi\tau}{w}\right)\cos(\pi\tau)\,\mathrm{d}\tau = \frac{2w}{\pi}\lambda(w)\sum_j y_jc_j\,,
\label{eq:I_ZY_exact}
\end{equation}
where we defined $y_j = (-1)^{j-1}s_j$ and $c_j = \cos(\pi t_j)$, and used that the term proportional to $\sin(\pi t_j)\sin(\pi\tau)$ is odd in $\tau$ and integrates to zero. The coefficient $\lambda(w)$ is given by
\begin{equation}
    \lambda(w) = \frac{\cos(\pi w/2)}{1-w^2}\,.
\label{eq:lambda_coeff}
\end{equation}
For $w \rightarrow 0$ we have $\lambda \rightarrow 1$, so that $\mathcal{I}_{ZY} \approx (2w/\pi)\sum_j y_jc_j$ vanishes linearly with the pulse width, recovering the result for instantaneous pulses. Since the infidelity in Eq.~\eqref{eq:infidelity_SE} is quadratic in the coefficients, the linear scaling of $\mathcal{I}_{ZY}$ with $w$ implies that the residual infidelity due to finite pulse width is $\mathcal{O}(w^2)$ unless $\sum_j y_j c_j = 0$. To evaluate $\mathcal{I}_{YZ} = \int_0^1 \sin(\vartheta)\cos(\varphi)\,\mathrm{d}t$ for pulses of finite width, we split the integration domain into the free-evolution intervals between pulses and the pulse windows $[t_j - w/2,\, t_j + w/2]$, and write $\mathcal{I}_{YZ} = \mathcal{I}_{YZ}^{\rm free} + \mathcal{I}_{YZ}^{\rm pulse}$. We assume that the pulses do not overlap and lie entirely within the sequence, i.e., $t_{j+1} - t_j \geq w$, $t_1 \geq w/2$, and $t_N \leq 1 - w/2$. For the free evolution, $\varphi = j\pi$ modulo $2\pi$ between the $j$-th and $(j+1)$-th pulse, so that $\cos(\varphi) = (-1)^j$ in this interval. Denoting the boundaries of the $j$-th free interval by $a_j$ and $b_j$, with $a_0 = 0$, $a_j = t_j + w/2$, $b_j = t_{j+1} - w/2$, and $b_N = 1$, we obtain
\begin{equation}
    \mathcal{I}_{YZ}^{\rm free} = \sum_{j=0}^{N} (-1)^j \left[G(b_j) - G(a_j)\right] = 2G\left(\tfrac{1}{2}\right) + \sum_{j=1}^{N} (-1)^{j-1}\left[G\left(t_j - \tfrac{w}{2}\right) + G\left(t_j + \tfrac{w}{2}\right)\right]\, ,
\label{eq:I_YZ_free_1}
\end{equation}
where we used $G(0) = 0$, $G(1) = 2G(1/2)$, and that $N$ is even. From Eq.~\eqref{eq:G_1} it follows that $G(t_j - w/2) + G(t_j + w/2) = 2G(t_j) + (2/\pi)\,c_j\left[1 - \cos(\pi w/2)\right]$, and therefore
\begin{equation}
    \mathcal{I}_{YZ}^{\rm free} = 2G\left(\tfrac{1}{2}\right) + 2\sum_{j=1}^{N}(-1)^{j-1}G(t_j) + \frac{2}{\pi}\left[1 - \cos\left(\frac{\pi w}{2}\right)\right]\sum_{j=1}^{N}(-1)^{j-1}c_j\, .
\label{eq:I_YZ_free_2}
\end{equation} During the $j$-th pulse, $\cos(\varphi) = (-1)^j\sin(\pi\tau/w)$ and $\sin(\vartheta) = \sin(\pi t_j)\cos(\pi\tau) + c_j\sin(\pi\tau)$. The first term is even in $\tau$ and does not contribute, since $\cos(\varphi)$ is odd. We thus find
\begin{equation}
    \mathcal{I}_{YZ}^{\rm pulse} = \sum_{j=1}^{N} (-1)^j c_j \int_{-w/2}^{w/2} \sin(\pi\tau)\sin\left(\frac{\pi\tau}{w}\right)\mathrm{d}\tau = -\frac{2w^2}{\pi}\,\lambda(w)\sum_{j=1}^{N}(-1)^{j-1}c_j\, ,
\label{eq:I_YZ_pulse}
\end{equation}
with $\lambda(w)$ given by Eq.~\eqref{eq:lambda_coeff}. Combining Eqs.~\eqref{eq:I_YZ_free_2} and \eqref{eq:I_YZ_pulse}, and using $G(t_j) = (1 - c_j)/\pi$, $\sum_{j=1}^{N}(-1)^{j-1} = 0$ for even $N$, and $\cos(\pi w/2) + w^2\lambda(w) = \lambda(w)$, we arrive at
\begin{equation}
    \mathcal{I}_{YZ} = \frac{2}{\pi}\left[1 - \lambda(w)\sum_{j=1}^{N}(-1)^{j-1}c_j\right].
\label{eq:I_YZ_exact}
\end{equation}
In the limit $w \rightarrow 0$, $\lambda \rightarrow 1$ and Eq.~\eqref{eq:I_YZ_exact} reduces to the result for instantaneous pulses, Eqs.~\eqref{eq:N_2_conditions}--\eqref{eq:N_6_conditions}. For small $w$, $1 - \lambda(w) = (\pi^2/8 - 1)\,w^2 + \mathcal{O}(w^4)$, so the finite width modifies $\mathcal{I}_{YZ}$ only at order $w^2$, in contrast to the linear scaling of $\mathcal{I}_{ZY}$. Note also that $\mathcal{I}_{YZ}$ is independent of the pulse signs $s_j$.\\
\indent We can now use $\mathcal{I}_{Z}^{(N)}$ from Eqs.~\eqref{eq:N_2_conditions}--\eqref{eq:N_6_conditions}, together with Eqs.~\eqref{eq:I_YZ_exact} and \eqref{eq:I_ZY_exact}, to determine the number of pulses and the times $t_j$ required for a finite width $w$. We have already shown in Eq.~\eqref{eq:N_2_solution} that dephasing and crosstalk suppression cannot be achieved for $N=2$. For $N=4$, the palindromic symmetry $s = (s_1, s_2, s_2, s_1)$ allows two sign sequences up to a global sign, which we denote by the signs of the first two pulses, namely $S_1 = (+,+)$ and $S_2 = (+,-)$. Using $c_3 = -c_2$ and $c_4 = -c_1$, Eq.~\eqref{eq:I_ZY_exact} gives
\begin{equation}
    \mathcal{I}_{ZY}^{S_1} = \frac{4w}{\pi}\lambda(w)\left(c_1 - c_2\right)\, , \qquad\qquad \mathcal{I}_{ZY}^{S_2} = \frac{4w}{\pi}\lambda(w)\left(c_1 + c_2\right)\, .
\label{eq:N_4_IZY}
\end{equation}
Since $\mathcal{I}_{YZ}$ does not depend on the pulse signs, it takes the same form for both sequences. Using $\sum_{j}(-1)^{j-1}c_j = 2(c_1 - c_2)$, Eq.~\eqref{eq:I_YZ_exact} gives
\begin{equation}
    \mathcal{I}_{YZ}^{S_1} = \mathcal{I}_{YZ}^{S_2} = \frac{2}{\pi}\left[1 - 2\lambda(w)\left(c_1 - c_2\right)\right],
\label{eq:N_4_IYZ}
\end{equation}
which vanishes for $c_1 - c_2 = 1/[2\lambda(w)]$. Combined with $t_2 = t_1 + 1/4$ from $\mathcal{I}_{Z}^{(4)} = 0$, this yields
\begin{equation}
    t_1(w) = \frac{1}{\pi}\arcsin\left[\frac{1}{4\sin(\pi/8)\,\lambda(w)}\right] - \frac{1}{8}\, , \qquad\qquad t_2(w) = t_1(w) + \frac{1}{4}\, ,
\label{eq:N_4_solutions_finite}
\end{equation}
which reduces to Eq.~\eqref{eq:N_4_solutions} for $w \rightarrow 0$. Inserting the solution $c_1 - c_2 = 1/[2\lambda(w)]$ into Eq.~\eqref{eq:N_4_IZY}, and using $c_1 + c_2 = 2\cos(\pi/8)\cos(\pi t_1 + \pi/8)$ together with Eq.~\eqref{eq:N_4_solutions_finite}, we obtain for the two sign patterns
\begin{equation}
    \mathcal{I}_{ZY}^{S_1} = \frac{2w}{\pi}\, , \qquad\qquad \mathcal{I}_{ZY}^{S_2} = \frac{8w}{\pi}\cos\left(\frac{\pi}{8}\right)\lambda(w)\sqrt{1 - \frac{1}{\left[4\sin(\pi/8)\,\lambda(w)\right]^2}}\, .
\label{eq:N_4_IZY_residual}
\end{equation}
The first is independent of the pulse times and cannot be cancelled, while the second is strictly positive, since both $c_1$ and $c_2$ are positive for $0 < t_1 < t_2 < 1/2$. Hence, for $N=4$ pulses of finite width, both sign patterns leave a residual crosstalk coefficient that grows linearly with $w$, and the crosstalk cannot be suppressed to first order. This motivates the use of $N=6$ pulses, where the additional degree of freedom can be used to cancel $\mathcal{I}_{ZY}$. For $N=6$, the pulses are centered around $t_1$, $t_2$, $t_3$, $1-t_3$, $1-t_2$, $1-t_1$, and the symmetry implies $c_4 = -c_3$, $c_5 = -c_2$, and $c_6 = -c_1$. As shown above, $\mathcal{I}_{Z}$ and $\mathcal{I}_{YZ}$ do not depend on the pulse signs, and therefore impose the same conditions on the times for every sign sequence. Using Eq.~\eqref{eq:N_6_conditions} and Eq.~\eqref{eq:I_YZ_exact} with $\sum_j(-1)^{j-1}c_j = 2(c_1 - c_2 + c_3)$, we obtain
\begin{equation}
    \mathcal{I}_{Z}^{(6)} = 4(t_1 - t_2 + t_3) - 1\, , \qquad\qquad \mathcal{I}_{YZ}^{(6)} = \frac{2}{\pi}\left[1 - 2\lambda(w)\left(c_1 - c_2 + c_3\right)\right],
\label{eq:N_6_IZ_IYZ}
\end{equation}
which vanish for
\begin{equation}
    t_1 - t_2 + t_3 = \frac{1}{4}\, , \qquad\qquad c_1 - c_2 + c_3 = \frac{1}{2\lambda(w)}\, .
\label{eq:N_6_time_conditions}
\end{equation}
The palindromic symmetry $s = (s_1, s_2, s_3, s_3, s_2, s_1)$ allows four sign sequences up to a global sign, namely $S_1 = (+,+,+)$, $S_2 = (+,-,+)$, $S_3 = (+,-,-)$, and $S_4 = (+,+,-)$. Equation~\eqref{eq:I_ZY_exact} then gives
\begin{equation}
    \mathcal{I}_{ZY}^{(6)} = \frac{4w}{\pi}\lambda(w)\left(s_1 c_1 - s_2 c_2 + s_3 c_3\right)\,.
\label{eq:N_6_IZY}
\end{equation}
Since $0 < t_1 < t_2 < t_3 < 1/2$, we have $c_1 > c_2 > c_3 > 0$, and therefore $\mathcal{I}_{ZY}^{(6)}$ is strictly positive for $S_1$, $S_2$, and $S_3$. Only $S_4$ allows $\mathcal{I}_{ZY}^{(6)} = 0$, which requires $c_1 = c_2 + c_3$. Together with Eq.~\eqref{eq:N_6_time_conditions}, this gives $c_3 = c_1 - c_2 = 1/[4\lambda(w)]$, and the three conditions are solved by
\begin{equation}
    t_3 = \frac{1}{\pi}\arccos\left[\frac{1}{4\lambda(w)}\right], \qquad\qquad t_1 = \frac{1}{\pi}\arcsin\left[\frac{1}{8\lambda(w)\sin(\pi\delta/2)}\right] - \frac{\delta}{2}\, , \qquad\qquad t_2 = t_1 + \delta\, ,
\label{eq:N_6_solutions}
\end{equation}
with $\delta = t_3 - 1/4$.\\
\indent Finally, for normalized gate time $T=1$, beamsplitter drive strength $g=\pi$, and $\Omega_{\rm SE}= 25$ [the same as the maximum amplitude in Fig.~\ref{fig:Crosstalk_and_dephasing}(b)], which gives $w = \pi/25 \simeq 0.1257$ and $\lambda = 0.9963$, Eq.~\eqref{eq:N_6_solutions} yields
\begin{equation}
    t_1 \simeq 0.0738\, , \qquad\qquad t_2 \simeq 0.2431\, , \qquad\qquad t_3 \simeq 0.4193\, .
\label{eq:N_6_numerical}
\end{equation}
\begin{figure}[t!]
\includegraphics[width=1\linewidth,keepaspectratio]{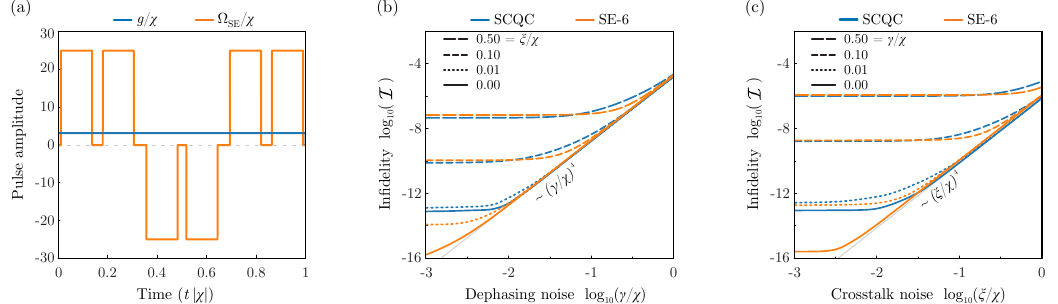}
    \caption{\label{fig:SE_step_2_results} \Filippos{Gate design and robustness tests for step (ii) using spin-echo pulses. (a) Control fields for the BS (blue) and the ancilla transmon (orange) versus time, in units of $1/|\chi|$, for the $X_1\otimes Z_2$ gate. The BS square pulse of amplitude $g/\chi = \pi$ implements the inter-cavity photon swap. The transmon drive consists of six spin-echo (SE-6) pulses of amplitude $\Omega_{\rm SE}/\chi = 25$, obtained from the analysis in Sec.~\ref{sec:spin_echo_X1Z2}, which suppresses dephasing and crosstalk to first order and implements $I_2$ up to a virtual-$Z$ rotation. (b) Gate infidelity versus $Z_2$ dephasing noise $\gamma/\chi$ for several values of the $ZZ$ crosstalk $\xi/\chi$ (line styles). (c) Gate infidelity versus $ZZ$ crosstalk $\xi/\chi$ for several values of the $Z_2$ dephasing $\gamma/\chi$ (line styles). In (b) and (c), the infidelity of the SE-6 gate (orange) is shown alongside that of the SCQC-designed gate from the main text [see Fig.~\ref{fig:Crosstalk_and_dephasing}] (blue). The gray lines indicate the $(\gamma/\chi)^4$ and $(\xi/\chi)^4$ scaling expected for first-order suppression.}}
\end{figure}
\indent The control fields, together with their robustness against dephasing and crosstalk, are shown in Fig.~\ref{fig:SE_step_2_results}. For both noise sources, the infidelity of the SE-6 gate scales as $\mathcal{O}(\gamma^4 + \xi^4)$ confirming the expected first-order cancellation of dephasing and crosstalk. In Fig.~\ref{fig:SE_step_2_results}(b), we observe that in the strong dephasing regime, $\gamma/\chi \geq 0.01$,  the SE-6 and SCQC gates perform essentially identically following the $(\gamma/\chi)^4$ slope. This behavior is also observed in the strong crosstalk regime, $\xi/\chi \geq 0.01$, as shown in Fig.~\ref{fig:SE_step_2_results}(c). For weaker noise, the SE-6 gate reaches lower infidelities, since its conditions are satisfied exactly by the analytic solution of Eq.~\eqref{eq:N_6_solutions}, whereas the SCQC pulses are smooth, time-dependent fields optimized to satisfy Eq.~\eqref{eq:crosstalk_condition} over a finite number of parameters, which leaves a small residual infidelity. However, this advantage is largely of theoretical interest since in both cases the gate infidelities are close to machine precision. Also, the SE-6 gate relies on ideal square pulses, whose sharp edges imply an unbounded bandwidth that cannot be realized in practice. Smoothing the pulse edges, as required by realistic control hardware, is expected to raise the residual infidelity floor and affect the performance of the SE approach in total. In contrast, the SCQC pulses are already smooth and therefore experimentally friendly, while achieving the same robustness as the SE-6 gate in the experimentally relevant range of noise strengths.
\subsection{Hybrid operations}
\noindent In the hybrid joint parity operation, the transmon control fields in steps (i) and (iii) are the SCQC-designed pulses of the main text [see Fig.~\ref{fig:Pulse_and_robustness}], while in step (ii) the control fields are replaced by a square pulse on the BS and the SE-6 sequence derived in Sec.~\ref{sec:spin_echo_X1Z2} and shown in Fig.~\ref{fig:SE_step_2_results}(a). The resulting false assignment probabilities and gate infidelities are shown in Fig.~\ref{fig:Parity_and_ZZgate}, together with those of the full SCQC gate design discussed in the main text.\\
\indent Overall, the hybrid and full SCQC operations perform very similarly, and both clearly outperform the non-robust pulses. This is expected, since step (ii) is short compared with steps (i) and (iii), so that replacing only the control fields in step (ii) has a small effect on the total accumulated error. The false negative probability [black lines in Fig.~\ref{fig:Parity_and_ZZgate}(a)] is essentially identical for both operations, because the states $|00\rangle$ and $|11\rangle$ are not affected by the $ZZ$ crosstalk, while both pulses suppress transmon dephasing to the same order, and hence the choice of pulse in step (ii) does not matter. For the false positive probability, the hybrid operation performs slightly better at weak dephasing and zero crosstalk, consistent with the lower residual infidelity of the SE-6 sequence observed in Fig.~\ref{fig:SE_step_2_results}. Importantly, the false positive probability of the hybrid operation still exhibits second-order suppression of dephasing, scaling as $(\gamma/\chi)^6$, which is a property of the geometrically designed pulse used in steps (i) and (iii). Apart from this regime, the full SCQC operation performs slightly better than the hybrid one for the false assignment probabilities, the joint parity gate, and the logical entangling gate in Fig.~\ref{fig:Parity_and_ZZgate}, as seen from the marginally lower plateaus at finite crosstalk.
}
\section{False positive and negative erasure checks \label{sec:false positive}}
\noindent The average gate fidelity is a faithful metric for quantifying the quality of the implemented operation. However, in the case of the joint parity operation, we are mainly interested in detecting whether we are in or out of the codespace with high accuracy. In other words, we are interested in the performance only for a subset of initial and final states. To that end, in this section we focus on two types of false assignments, the false negative (FN) where the dual-rail qubit has leaked out of the codespace and the parity check fails to detect this event, and the false positive (FP) where we mistakenly reset the dual-rail qubit while it is still in the codespace.\\
\indent Following the joint parity protocol shown in Fig.~\ref{fig:circuits}(b), in the noise-free case, the ancilla qubit is initialized in the ground state, $\ket{\psi_{\rm i}} = \ket{\textsl{g}}$. During the protocol, it is measured either in the ground state, $\ket{\psi_{\rm f}} = \ket{\textsl{g}}$, or in the excited state, $\ket{\psi_{\rm f}} = \ket{f}$, indicating that the dual-rail state has remained within or leaked out of the codespace, respectively. However, in the presence of noise the ancilla can end up in the wrong state, which will lead to a false assignment as described above. The probability of a FN event is given as the probability of measuring the ancilla in the ground state given that the dual-rail qubit has leaked out of the codespace, $\ket{\phi}\in\mathcal{C}^{\bot}=\{\ket{00}, \ket{11}\}$, while the probability of a FP event is given as the probability of measuring the ancilla in the excited state given that the dual-rail qubit is in the codespace, $\ket{\phi}\in\mathcal{C}=\{\ket{01}, \ket{10}\}$. These two probabilities are calculated as follows
\begin{subequations}
\begin{equation}
    P({\rm FN}) = P(|\braket{\textsl{g}|\psi_{\rm f}}|^2 \mid \ket{\phi}\in \mathcal{C}^\bot)=  \bra{\textsl{g}}{\rm tr}_{\phi}\big[\mathcal{U}_{\rm JP}(\rho_{\phi}\otimes \ket{\textsl{g}}\bra{\textsl{g}})\mathcal{U}^{\dagger}_{\rm JP}\big] \ket{\textsl{g}}\, ,
\label{eq:false_negative}
\end{equation}
\begin{equation}
    P{\rm(FP)} = P(|\braket{f|\psi_{\rm f}}|^2 \mid \ket{\phi}\in \mathcal{C})= \bra{f}{\rm tr}_{\phi}\big[\mathcal{U}_{\rm JP}(\rho_{\phi}\otimes  \ket{\textsl{g}}\bra{\textsl{g}})\mathcal{U}^{\dagger}_{\rm JP}\big] \ket{f}\, ,
\label{eq:false_positive}
\end{equation}
\end{subequations}
 where $\mathcal{U}_{\rm JP} = (I_1\otimes Y_{-\frac{\pi}{2}})\, U_{\rm JP}\, (I_1\otimes Y_{-\frac{\pi}{2}})$, and $\rho_{\phi}\otimes\ket{\textsl{g}}\bra{\textsl{g}}$ is the initial state of the system, with $\rho_{\phi}$ the density matrix of the dual-rail qubit. For FN misassignment, $\rho_{\phi}$ is one of the leakage states, $\ket{00}\bra{00}$ or $\ket{11}\bra{11}$. For FP misassignment, $\rho_{\phi} = \ket{\phi}\bra{\phi}$, where $\ket{\phi}$ is an arbitrary state in the dual-rail subspace spanned by $\{\ket{01},\ket{10}\}$. In Fig.~\ref{fig:False_negative_main}, we plot the above probabilities in the presence of ancilla dephasing noise, for the non-robust and robust joint parity operations. As expected, in the case of non-robust joint parity checks both probabilities scale with $\propto\gamma^2$, showing zero protection against noise. In contrast, for the robust joint parity check the FN probability scales like the 4th power of the noise strength, $P({\rm FN})\propto \gamma^4$, while the FP probability has sextic sensitivity to noise, $P({\rm FP}) \propto \gamma^6$, revealing thus first- and second-order dephasing noise suppression, respectively. The first-order suppression in FN events are expected due to the closed curves, while the additional order of suppression shown in the case of FP is due to the $X_1\otimes Z_2$ step in our protocol, as explained in detail in Sec.~\ref{sec:second_order_cancellation}. \\

\indent At this point we should mention that the effect and the importance of each misassignment is different. ``False positive" errors are the most frequent type of misassignment, and they increase the erasure rate \cite{Teoh2023, Chou2024} in the dual-rail qubit. However, they do not lead to logical errors and thus are not too harmful. In contrast, FN errors occur less frequently, as they arise from a combination of failure mechanisms and therefore constitute second-order events. An FN error occurs when the dual-rail qubit leaks out of the codespace (with probability $p_{\rm erasure}$) and the erasure check operation fails to detect it. However, this undetected leakage event can lead to logical errors, which are among the most damaging errors in a stabilizer code, and thus it is important to keep the fraction of leaked qubits small \cite{Chou2024}. That said, the non-symmetric cancellation of ancilla dephasing noise shown in Fig.~\ref{fig:False_negative_main} works in our favor by suppressing the erasure rate to second order and the already rare false negative outcomes to first order.
\begin{figure}[t!]
\includegraphics[width=1\linewidth,keepaspectratio]{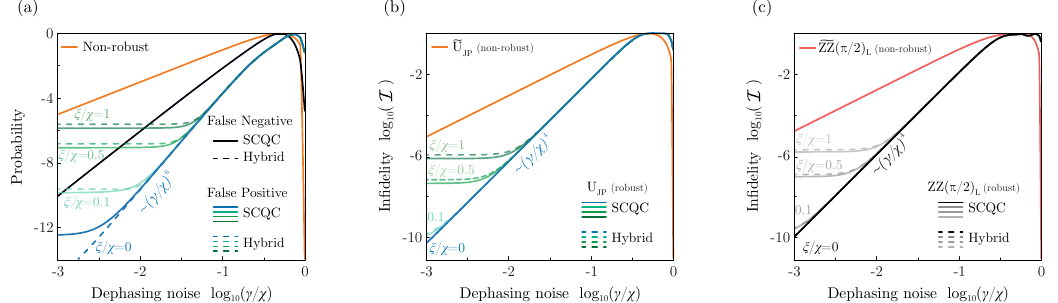}
\caption{\label{fig:Parity_and_ZZgate}\Filippos{Robustness of the erasure check and the logical gates against ancilla dephasing $\gamma/\chi$ for different values of the $ZZ$ crosstalk $\xi/\chi$. Solid lines correspond to gates designed entirely with SCQC, and dashed lines to the hybrid scheme, in which step (ii) is replaced by the SE-6 sequence of Fig.~\ref{fig:SE_step_2_results}. Darker shades correspond to larger values of $\xi/\chi$. (a) False erasure assignment probabilities. For the robust erasure check, the false negative probability (black) scales as $(\gamma/\chi)^4$, while the false positive probability (blue-green) scales as $(\gamma/\chi)^6$ and saturates when the crosstalk becomes the dominant noise source. For the non-robust erasure check (orange), the false negative and false positive probabilities coincide and scale as $(\gamma/\chi)^2$. The false negative probability does not include the probability of leaking out of the codespace, $p_{\rm erasure}$. (b) Gate infidelity $\mathcal{I} = 1 - \mathcal{F}$ of the robust $U_{\rm JP}$ (blue-green) and non-robust $\widetilde{U}_{\rm JP}$ (orange) joint parity gates. (c) Gate infidelity of the robust $ZZ(\pi/2)_{\rm L}$ (black-gray) and non-robust $\widetilde{ZZ}(\pi/2)_{\rm L}$ (red) logical entangling gates. In (b) and (c), the infidelity of the non-robust gates scales as $(\gamma/\chi)^2$, while that of the robust gates scales as $(\gamma/\chi)^4$ and reaches a plateau set by $\xi/\chi$. Here $\xi$ is the unwanted dispersive coupling in step (ii) of the joint parity sequence, and $\chi$ is the coupling in steps (i) and (iii).}}
\end{figure}
\section{Construction of \texorpdfstring{$ZZ(\theta)_{\rm L}$}{TEXT} from the joint parity unitary \label{sec:logical ZZ }}
\noindent Here we follow an analysis similar to Ref.~\cite{Teoh2023} and show how to construct a dynamically corrected $ZZ(\theta)_{\rm L}$ entangling gate from the already robust joint parity unitary, $U_{\rm JP}$, derived in Sec.~\ref{sec:three_step parity check}. We show that the following sequence of unitaries
\begin{equation}
    Y_2\left(\frac{\pi}{2}\right)U_{\rm JP}X_2(-\theta)U_{\rm JP}Y_2\left(-\frac{\pi}{2}\right) = e^{i\frac{\pi}{4}Y_2} U_{\rm JP} e^{-i\frac{\theta}{2}X_2}U_{\rm JP}e^{-i\frac{\pi}{4}Y_2}
\label{eq:ZZ_sequence}
\end{equation}
realizes the unitary $ZZ(\theta)$ where
\begin{equation}
    ZZ(\theta) = e^{-i\frac{\theta}{2}Z_1\otimes Z_2} \equiv e^{-i\frac{\theta}{2}}\begin{pmatrix}
                    1 & \phantom{1} & \phantom{1} & \phantom{1}\\
                    \phantom{1} & e^{i\theta} & \phantom{1} & \phantom{1} \\
                    \phantom{1} & \phantom{1} & e^{i\theta} & \phantom{1}\\
                    \phantom{1} & \phantom{1} & \phantom{1} & 1
                \end{pmatrix} 
\label{eq:ZZ_theta}
\end{equation}
acting on the joint-cavity subspace $\{\ket{00},\ket{01},\ket{10},\ket{11}\}$. When each of the cavities belongs to a rail in a distinct DR qubit, implementing a physical $ZZ(\theta)$ gate on them amounts to performing a logical $ZZ_{\rm L}(\theta)$ operation on the logical qubits. To demonstrate why this construction works, we will start with the example state $\ket{\Psi} = (\ket{00} + \ket{01} + \ket{10} + \ket{11})\otimes \ket{\textsl{g}}$ (ignoring normalization) and apply the unitaries shown in Eq. \eqref{eq:ZZ_sequence} sequentially. \\
\indent 1. Operation $e^{-i\frac{\pi}{4}Y_2}$ puts the ancilla in the $\ket{+} = \frac{\ket{\textsl{g}} + \ket{f}}{\sqrt{2}}$ state
\begin{equation}
    (\ket{00} + \ket{01} + \ket{10} + \ket{11})\otimes \ket{+} .
\end{equation}
\indent 2. The joint parity unitary, $U_{\rm JP}$, introduces a $\pi$ phase on the cavity states $\ket{01}$ and $\ket{10}$ if the ancilla is in the $\ket{\textsl{g}}$ state. It also swaps the photons in the cavities because of the $X_1$ operator in Eq. \eqref{eq:Joint_parity_final} but this does not affect the resulted state. Hence, upon the second unitary the state can be written as
\begin{equation}
    (\ket{00} + \ket{11})\otimes\ket{+} - (\ket{01} + \ket{10})\otimes\ket{-}\, ,
\end{equation}
where $\ket{-} = \frac{\ket{\textsl{g}} - \ket{f}}{\sqrt{2}}$.\\
\indent 3. The ancilla rotation performed by $e^{-i\frac{\theta}{2}X_2}$ is where the phase is imprinted on the cavities. This unitary transforms $\ket{\pm}$ states to $e^{\mp i \frac{\theta}{2}}\ket{\pm}$, thus the total state becomes
\begin{equation}
    e^{-i\frac{\theta}{2}}(\ket{00} + \ket{11})\otimes\ket{+} - e^{+i\frac{\theta}{2}}(\ket{01} + \ket{10})\otimes\ket{-}\, .
\end{equation}
\indent 4. Next, we disentangle the ancilla from the cavities by applying another joint parity unitary. The $X_1$ operation, hidden in the joint parity unitary, cancels the one implemented in the first application of $U_{\rm JP}$ in the second step. The state reads 
\begin{equation}
    \left[e^{-i\frac{\theta}{2}}(\ket{00} + \ket{11}) + e^{+i\frac{\theta}{2}}(\ket{01} + \ket{10})\right] \otimes\ket{+}\, .
\end{equation}
\indent 5. Finally, the $e^{i\frac{\pi}{4}Y_2}$ puts the ancilla back to the ground state $\ket{\textsl{g}}$, Also, by measuring the ancilla state at the end, it error-detects the gate. 
\begin{equation}
    \left[e^{-i\frac{\theta}{2}}(\ket{00} + \ket{11}) + e^{+i\frac{\theta}{2}}(\ket{01} + \ket{10})\right] \otimes\ket{\textsl{g}}\, .
\label{eq:ZZ_operation_final}
\end{equation}
From Eq. \eqref{eq:ZZ_operation_final}, the unitary applied to the cavities is
\begin{equation}
    \bra{\textsl{g}}e^{i\frac{\pi}{4}Y_2} U_{\rm JP} e^{-i\frac{\theta}{2}X_2}U_{\rm JP}e^{-i\frac{\pi}{4}Y_2}\ket{\textsl{g}} = e^{-i\frac{\theta}{2}}  \begin{pmatrix}
                        1 & \phantom{1} & \phantom{1} & \phantom{1}\\
                        \phantom{1} & e^{i\theta}  & \phantom{1} & \phantom{1} \\
                        \phantom{1} & \phantom{1} & e^{i\theta}  & \phantom{1}\\
                        \phantom{1} & \phantom{1} & \phantom{1} & 1 
                    \end{pmatrix}  \, ,
\end{equation}
which is \Filippos{exactly} Eq. \eqref{eq:ZZ_theta}. Furthermore, $ZZ(\theta= \pi/2)$ is locally equivalent to \textsc{cz} and \textsc{cnot} and thus a perfect entangler. Recall that the joint parity operator is robust to first order in the ancilla dephasing noise, and the single-qubit rotations on the ancilla qubit can be made robust using SCQC or simply by using \texttt{qurveros} \cite{Piliouras2026}. Figure~\ref{fig:Parity_and_ZZgate}(c) depicts the performance of the non-robust (red) and robust (gray scale) $ZZ(\pi/2)_{\rm L}$ against ancilla-dephasing noise strength. Our robust logical gate suppresses noise to first order and has higher fidelity (lower infidelity) for all the noise strengths compared to the non-robust counterpart. Furthermore, our gate performs better even in the case where there is no tunability over the dispersive coupling, $\xi/\chi = 1$, (light gray). 
\section{Additional results \label{sec:additional_results}}
\begin{figure}[t]
\includegraphics[width=1\linewidth,keepaspectratio]{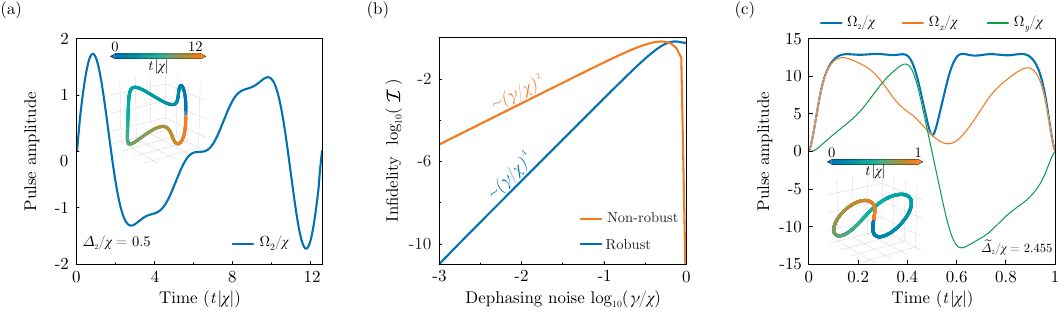}
    \caption{\label{fig:Additional result} Additional waveforms. (a)-(b) Gate design for steps (i) and (iii) derived in Sec.~\ref{sec:three_step parity check}. (a) Control field $\Omega_2(t)$ on the transmon ancilla qubit for the $ZZ(\pi/2)$ gate obtained from the curvature and the torsion of the curve shown in the inset. (b) Gate infidelity $(\mathcal{I} = 1-\mathcal{F})$ vs dephasing noise strength $(\gamma/\chi)$ showing the robustness against quasi-static dephasing noise to leading-order for the $ZZ(\pi/2)$ gate (blue). For comparison, the infidelity of the non-robust square pulse with amplitude $\Omega/\chi = 5.5\pi$ is also shown (orange). (c) Control fields, obtained through BARQ, for the ancilla qubit for the $Z_2$ gate needed in step (ii) derived in Sec.~\ref{sec:three_step parity check}, only in the case where $\xi=0$. $X_1$ gate is not affected and can be implemented with the Gaussian pulse shown in Fig.~\ref{fig:Crosstalk_and_dephasing}(a).}
\end{figure}

\noindent In this section, we present additional results that complement those in the main text and further demonstrate the versatility of our approach. In particular, we show that our method can generate a variety of control waveforms that can be adapted to the requirements and constraints of different dual-rail implementations.\\
\indent As it is mentioned in the main text, we can design a curve for the noise-free part of $H_{+}$ with target gate $U_g = Z(\pi/2)$ using the family of closed curves that satisfies the criteria of constant torsion and smooth curvature \cite{Calini1996}. Here, as an ansatz, we set $p = 0.90890$, which is a closed curve with constant torsion $\tau = 2.86725\cdot10^{-3} $ \cite {Calini1996}, that leads to a $R_z(2\pi)$ rotation. Similar to the main text, we adjust the torsion $\tau$ and the Fourier components of the curvature $\kappa(t)$ until we obtain $U_{+}(T_g) = R_z(\pi/2)$ with infidelity $\mathcal{I} = 1-\mathcal{F} = 10^{-14}$. Hamiltonian $H_{-}$ evolves with the same curvature $\kappa(t)$ but negative torsion, and so the gate in that block is $U_{-}(T_g) = XR_z(\pi/2)X = R_z(-\pi/2)$, see Sec.~\ref{subsec:negative_torsion}. Combining these two pieces together, we realize the entangling operation $U_{0}^{ZZ}(T_g) = \ket{0}\bra{0}\otimes U_{-} +\ket{1}\bra{1}\otimes U_{+} = ZZ(\pi/2)$ with a smooth driving field given by $\Omega_2(t) = \kappa(t)$ and constant detuning $\Delta_2 = \chi/2$. Figure \ref{fig:Additional result}(a)-(b) depicts the corresponding curve and the experimentally friendly waveform along with the infidelity of the gate against the noise strength. The obtained waveform is shorter in gate time, $T_g = 12.6/|\chi| $, while it is steeper in both ends and reaches higher peak values compared to the waveform shown in Fig.~\ref{fig:Pulse_and_robustness}(a). The slope of the DCG infidelity is proportional to $\gamma^4$, demonstrating first-order dephasing cancellation due to our closed-curve design, while the non-robust gate infidelity follows the expected $\gamma^2$ dependence.\\
\indent In Sec.~\ref{sec:crosstalk_condition} we derive the geometric condition for DCG against quasi-static $ZZ$ crosstalk noise, while we also provide waveforms that satisfy this condition. In Fig.~\ref{fig:Parity_and_ZZgate} we show the effect of this unwanted interaction assuming finite tunability over dispersive coupling during step (ii) in our protocol, see Sec.~\ref{sec:three_step parity check}. However, the newly proposed linear inductive coupler (LINC) ~\cite{Maiti2025} makes the dispersive coupling tunable and can achieve $\xi/\chi \sim10^{-4}$, which is effectively zero. In this case, Eq.~\eqref{eq:crosstalk_condition} is redundant and does not need to be satisfied in step (ii) or our protocol. Hence, we can keep the Gaussian waveform for the dual-rail qubit and use BARQ \cite{Piliouras2026} for the ancilla to obtain control fields that are expected to be lower in amplitude and smoother. This expectation stems from the fact that we reduce the number of constraints in BARQ's cost function. For instance, when $\xi = 0$ one can use the control fields for the ancilla shown in Fig.~\ref{fig:Additional result}(c) and achieve the same performance. Notice, that this is not the only solution since BARQ can generate waveforms with zero detuning too.
\Filippos{
\section{Experimental Considerations \label{sec:exp_considerations}}
\noindent The pulse sequence is designed to be robust against quasi-static frequency errors on the ancillary transmon, which the closure condition Eq.~\eqref{eq:closed_curve_condition} suppresses to first order. This is the robustness plotted in Figs.~\ref{fig:Pulse_and_robustness}(b) and \ref{fig:Crosstalk_and_dephasing}(c) of the main text and Fig.~\ref{fig:Additional result}(b) of this Supplement. The pulses are also compatible with current hardware in terms of timing, amplitude, and bandwidth, as we discuss below.\\
\indent Timing errors are negligible by a wide margin. For $|\chi|/2\pi = 1$~MHz the gate times are $T_g \sim 0.16$--$2.5\ \upmu$s, whereas current transmon control electronics resolve gate durations down to a few nanoseconds and with typical sampling rates at $f_s = 2$~GHz, timing jitter falls more than two orders of magnitude below the timescales relevant to our pulses.\\
\indent Amplitude errors, which originate mainly in the amplification chain, are similarly not expected to be seen in our pulses. Our control amplitudes are up to about 25~MHz, well below the $>$100~MHz range at which transmon drives operate routinely. Furthermore, the envelopes are designed to be smooth, so neither amplifier nonlinearity nor bandwidth limitation is encountered. Should first-order robustness against multiplicative amplitude noise be desired, it can be imposed within the same framework by requiring the tangent curve to enclose vanishing signed area~\cite{Nelson2023}.\\
\indent Finally, the beam-splitter drive uses a Gaussian envelope, chosen as the smoothest available option, whose peak amplitude remains within the range reported in existing beam-splitter experiments \cite{deGraaf2025}.
}

%TC:endignore

\end{document}